\documentclass[aps,amsmath,amssymb,twocolumn,superscriptaddress]{revtex4}
\usepackage[]{graphicx}
\usepackage{color}
\usepackage{bbold}
\usepackage{braket}
\usepackage[colorlinks=true, pdfstartview=FitV, linkcolor=blue, citecolor=red, urlcolor=black]{hyperref}
\newcommand{\be}{\begin{equation}}
\newcommand{\ee}{\end{equation}}
\newcommand{\ben}{\begin{eqnarray}}
\newcommand{\een}{\end{eqnarray}}
\newcommand{\bes}{\begin{subequations}}
\newcommand{\ees}{\end{subequations}}
\def\bal#1\eal{\begin{align}#1\end{align}}
\usepackage{comment}

\newcommand{\nn}{\nonumber\\}
\newcommand{\bfi}{\begin{figure}}
\newcommand{\efi}{\end{figure}}
\newcommand{\bc}{\begin{center}}
\newcommand{\ec}{\end{center}}
\newcommand{\sech}{\mbox{sech}}

\newcommand{\sn}{\mbox{sn}}
\newcommand{\dn}{\mbox{dn}}

\begin{document}
    \title{Scalar-tensor representation to $f(R,T)-$brane with Gauss-Bonnet gravity}

\author{A.S. Lob\~ao}
    \affiliation{Escola T\'ecnica de Sa\'ude de Cajazeiras, Universidade Federal de Campina Grande, 58900-000 Cajazeiras, PB, Brazil}    
    \author{Jo\~ao Lu\'is Rosa}
    \affiliation{Institute of Physics, University of Tartu, W. Ostwaldi 1, 50411 Tartu, Estonia}
	\affiliation{University of Gda\'{n}sk, Jana Ba\.{z}y\'{n}skiego 8, 80-309 Gda\'{n}sk, Poland}
    \author{D. Bazeia}
    \affiliation{Departamento de F\'\i sica, Universidade Federal da Para\'\i ba, 58051-970 Jo\~ao Pessoa, PB, Brazil}

\begin{abstract}
In this paper, we generalize the analysis of the $f(R,T)-$brane via the inclusion of a term proportional to the Gauss-Bonnet invariant. We consider an action of the form $F(R,G,T)=f(R,T)+\alpha G$, where $T$ is the trace of the stress-energy tensor, $R$ is the Ricci scalar, and $\alpha$ is a real parameter that controls the contribution of the Gauss-Bonnet invariant $G$. We introduce the first-order formalism to obtain solutions for the source field of the brane in the special case where $f(R,T)=R+\beta T$ and illustrate its procedure with an application to the sine-Gordon model. We also investigate the general case of the $f(R,T)-$brane via the use of the scalar-tensor formalism, where we also use the first-order formalism to obtain solutions. Finally, we investigate the linear stability of the brane under tensor perturbations of the the modified Einstein's field equations. Our results indicate that the Gauss-Bonnet term may induce qualitatively different behaviors of the quantities on the brane, provided that its contribution is large enough.
\end{abstract}
\maketitle
\section{Introduction}

One of the most interesting issues in the current cosmological scenario is undoubtedly the nature of the constitution of the cosmos, especially the nature of dark matter and dark energy. Understanding why dark matter and dark energy are so widely present in the universe, being much more abundant than visible matter, is a great challenge \cite{Project:1998vns,Team:1998fmf,Team:2004lze,WMAP:2003elm}. It is known that the standard Einstein's theory of gravity is not able to explain such a discrepancy. In this sense, many researchers have tried to extend the cosmological model in order to clarify such questions. Such proposals are commonly known as modified gravity models \cite{Bamba:2012cp,Capozziello:2003tk,Bengochea:2008gz}.

The most widespread theory of modified gravity is the so-called $f(R)-$gravity, which includes a generic function of the scalar curvature in the Einstein-Hilbert action \cite{Buchdahl:1970, Starobinsky:1980te, Sotiriou:2008rp}. Many other generalizations have been presented with e.g. the inclusion of a quadratic invariant, such as the Gauss-Bonnet scalar defined by $G=R^2-4R_{ab}R^{ab}+R_{abcd}R^{abcd}$ \cite{Lanczos:1938,Lovelock:1971,Glavan:2019inb,Li:2007jm,Fronimos:2021ejo}, models with torsion \cite{Yoon:1999ie, Kranas:2018jdc, Pereira:2022cmu, Iosifidis:2021iuw,Ilyas:2022ieq} or even the coupling of the trace of the moment energy tensor with the scalar of curvature, generating the so-called $f(R,T)-$gravity \cite{Harko:2011kv,Cai:2015emx,Houndjo:2011tu,Alvarenga:2013syu}.

In addition to the investigation of cosmological scenarios in four dimensions, generalized models of gravity have also been used to address other contexts of interest, such as gravity in two dimensions in the presence of a dilation field \cite{Teitelboim:1983ux,Jackiw:1985,Zhong:2021voa} or braneworld models in five dimensions \cite{RS:1999,GW:1999,ST:1999, DWF:2000}. Particularly, in the context of braneworld models, modifying gravity can induce the emergence of an internal structure, in addition to changing the way the graviton interacts with the extra dimension, see e.g. \cite{Afonso:2007gc,Zhong:2010ae,HoffdaSilva:2011si,Bazeia:2013uva,Bazeia:2015owa,Cui:2018hqx,Dzhunushaliev:2019wvv}.

Recently, a dynamically equivalent scalar-tensor representation of the $f\left(R,T\right)$ theory of gravity, in which the arbitrary dependence of the function $f\left(R,T\right)$ in $R$ and $T$ is exchanged by two auxiliary scalar fields, thus reducing the order of the field equations from fourth to second order, has been proposed \cite{Rosa:2021teg,Rosa:2022cen} and immediately proven useful in the analysis of cosmological scenarios \cite{Goncalves:2022ggq,Goncalves:2021vci}. In recent works, models of the $f(R,T)-$brane have been investigated using this formalism \cite{Rosa:2021tei,Rosa:2021myu,Rosa:2022fhl,Bazeia:2022agk}. In general, obtaining solutions for the $f(R,T)-$brane is only possible in some simple cases, mainly in some low-order polynomial expansions of $R$ and $T$. However, the scalar-tensor representation allows for a detailed view of the problem without requiring us to make a specific choice for the form of the $f(R,T)$. In this context, it has been shown that the inclusion of modifications in the dynamics of the brane source field can cause interesting modifications in the behavior of the fields of the scalar-tensor representation. This investigation can also be generalized to include other modifications of interest as e.g. the inclusion of quadratic invariants such as the Gauss-Bonnet invariant, or also invariants that simulate a cubic gravity \cite{Oliva:2010eb,Myers:2010ru,Bueno:2016xff}. It is known that the Gauss-Bonnet invariant alone can induce changes in the brane structure, see e.g. the work described in Refs. \cite{Maeda:2007cb,Xu:2022ori,Bazeia:2015dna,Xu:2022gth}. The inclusion of this term together with the scalar-tensor representation can induce new behaviors.

In this paper, we investigate braneworld models in $F(R,G,T)-$gravity, i.e., via the inclusion of the Gauss-Bonnet invariant $G$ together with the trace of the stress-energy tensor $T$ and Ricci scalar. This proposal was first introduced in the cosmological scenario in \cite{Yousaf:2018lvl}. However, here we use the simpler form $F(R,G,T)=f(R,T)+\alpha G$, where $\alpha$ is a real parameter that controls the contribution of the Gauss-Bonnet invariant. We start the investigation in Sec. \ref{formaism}, where we present the main definitions and obtain the equations of motion that describe the brane model. In Sec. \ref{sec_sca_tens} we investigate the general case of $f(R,T)-$brane through the scalar-tensor representation. In Sec. \ref{stabiline} we obtain the conditions for the general model to be stable under tensor perturbations. We conclude the work in Sec. \ref{coments} with our final comments and remarks. Throughout this work Latin, indices $a, b, c, \cdots$, run from 0 to 4, while the Greek indices $\mu, \nu, \sigma, \cdots$, run from 0 to 3. We consider a system of geometrized units for which $c=1$, where $c$ is the speed of light, and the metric signature as $(+, -, -, -)$.

\section{Field Equations}\label{formaism}

In this work, we follow the idea presented in \cite{Yousaf:2018lvl} and consider a generalized Einstein-Gauss-Bonnet gravity in five dimensions that describes a $F(R,G,T)-$brane, where $R$ is the Ricci scalar, $G$ is the Gauss-Bonnet invariant and $T$ is the trace of stress-energy tensor. Moreover, for simplicity we consider a linear contribution of the Gauss-Bonnet invariant $G$, i.e., $F(R,G,T)=f(R,T)+\alpha G$, with $\alpha$ a real constant. This theory is described by an action of the form
\begin{equation}\label{actiongeo}
S=\frac1{2\kappa}\int_\Omega\!\!\sqrt{|g|}\,d^5x\big[f\left(R,T\right)+\alpha G\big]+S_m\left(\phi,g_{ab}\right),
\end{equation}
where $\kappa$ is a positive coupling constant, $\Omega$ is the five-dimensional spacetime manifold described by a set of coordinates $x^a$, and $g$ is the determinant of the metric tensor $g_{ab}$. Moreover, $S_m\left(\phi,g_{ab}\right)$ is the matter action that introduces the source field $\phi$ of the braneworld model. For the purpose of this work, we consider that $S_m$ is given by
\begin{equation}\label{lagrange}
S_m\left(\phi,g_{ab}\right)=-\int\!\!\sqrt{|g|}\,d^5x\left(\frac12 \nabla_a\phi\nabla^a\phi-V(\phi)\right).
\end{equation}

The modified field equations can be obtained by taking a variation of Eq. \eqref{actiongeo} with respect to the metric $g_{ab}$. These equations take the form
\be\label{EinsEq}
\begin{aligned}
&f_{R}R_{ab}+\big(g_{ab}\Box -\nabla_a\nabla_b\big)f_R-\frac12g_{ab}f\\
&+\alpha H_{ab}=\kappa T_{ab}-\big(T_{ab}+\Theta_{ab}\big)f_T,
\end{aligned}
\ee
where we defined $f_{R}\equiv \partial f/\partial R$, $f_{T}\equiv\partial f/\partial T$, $\Theta_{ab}= g^{cd}(\delta T_{cd}/\delta g^{ab})$, $\Box\equiv \nabla_a\nabla^a$ is the d'Alembert operator, with $\nabla_a$ denoting the covariant derivatives, and $H_{ab}$ is the Lovelock tensor defined in terms of the Riemann tensor and its contractions as
\be\label{hab}
\begin{aligned}
\!\!H_{ab}=&\,2RR_{ab}\!-\!4R_a{}^c R_{bc}\!+\!4R_{acdb}R^{cd}\!+\!2R_{acdl}R_{b}{}^{cdl}\\
\!\!&-\frac12g_{ab}\left(R^2-4R_{cd}R^{cd}+R_{cdmn}R^{cdmn}\right).
\end{aligned}
\ee

Taking the matter action for the source field as described in Eq. \eqref{lagrange}, the stress-energy tensor takes the form
\begin{equation}
T_{ab}=\nabla_a\phi\nabla_b\phi-\frac12g_{ab}\nabla_c\phi\nabla^c\phi+g_{ab}V(\phi).
\end{equation}
Consequently, the trace $T\equiv g^{ab}T_{ab}$ can be written in the form $T=-(3/2)\nabla_c\phi\nabla^c\phi+5V(\phi)$. Furthermore, the tensor $\Theta_{ab}$ is given by
\begin{equation}
\Theta_{ab}=-\frac52 \nabla_a\phi\nabla_b\phi+\frac12g_{ab}\nabla_c\phi\nabla^c\phi-g_{ab}V(\phi).
\end{equation}

The equation of motion for the source field $\phi$ is modified by the coupling of the trace $T$ with the geometry, taking the form
\ben
\nabla_a\nabla^a\phi+V_\phi =-\frac{3}{2\kappa}\nabla_a\big(f_T \nabla^a\phi\big)-\frac{5}{2\kappa}f_T V_\phi,
\een
where $V_\phi=dV/d\phi$. In this work, we restrict our analysis to models that generate kink-like solutions. For this purpose, we assume that the source field is static and depends only on the extra dimension $y$, i.e. $\phi=\phi(y)$. Furthermore, we consider the usual brane metric as
\begin{equation}\label{metricbrane}
    ds^2=e^{2A(y)}\eta_{\mu\nu}dx^\mu dx^\nu-dy^2\,,
\end{equation}
where $\eta_{\mu\nu}$ is the four-dimensional Minkowski metric and $A(y)$ is the warp function. For static configurations, the scalar functions $R$, $G$ and $T$ take the forms  $R=20A'^2+8A''$, $G=120 A'^4+96 A'^2 A''$, and $T=(3/2)\phi'^2+5V$, respectively, where a prime denotes derivatives with respect to $y$. Furthermore, the Kretschmann scalar $K$ defined by a contraction of the Riemann tensor in the form $K=R_{abcd}R^{abcd}$ can be written for static configurations as $K=40A^{\prime 4}+ 16A^{\prime\prime 2}+32A^{\prime 2}A^{\prime\prime}$. Under these assumptions, the equation of motion for the source field $\phi$ takes the form
\begin{eqnarray}\label{Eqmovphigeral}
\left(\!1\!+\!\frac{3f_T}{2 \kappa}\!\right)\left(\phi''\!+\!4A'\phi'\right)\!+\!\frac{3f_T'}{2\kappa} \phi'\!=\!\left(\!1\!+\!\frac{5f_T}{2\kappa}\!\right) \!V_\phi,
\end{eqnarray}
whereas the two non-vanish and independent components of the modified Einstein field equations in Eq. \eqref{EinsEq} are
\bes
\bal
&-3A''f_R+A'f_R'-f_R''-12\alpha A'^2 A''\!=\!\left(\!\kappa\!+\!\frac{3f_T}{2}\!\right)\phi'^2,\label{EqEins2.1geral}\\
&2\left(A'^2+A''\right)f_R-\frac14f-2A'f_R'-6\alpha A'^4\nn
&=-\frac14\left(\kappa+3f_T\right)\phi'^2\!+\!\frac12\kappa V.\label{EqEins2.2geral}
\eal
\ees

\subsection{Special case}\label{speccase}
\begin{figure*}[t!]
\centering
\includegraphics[width=7cm,trim={0cm 0cm 0 0cm},clip]{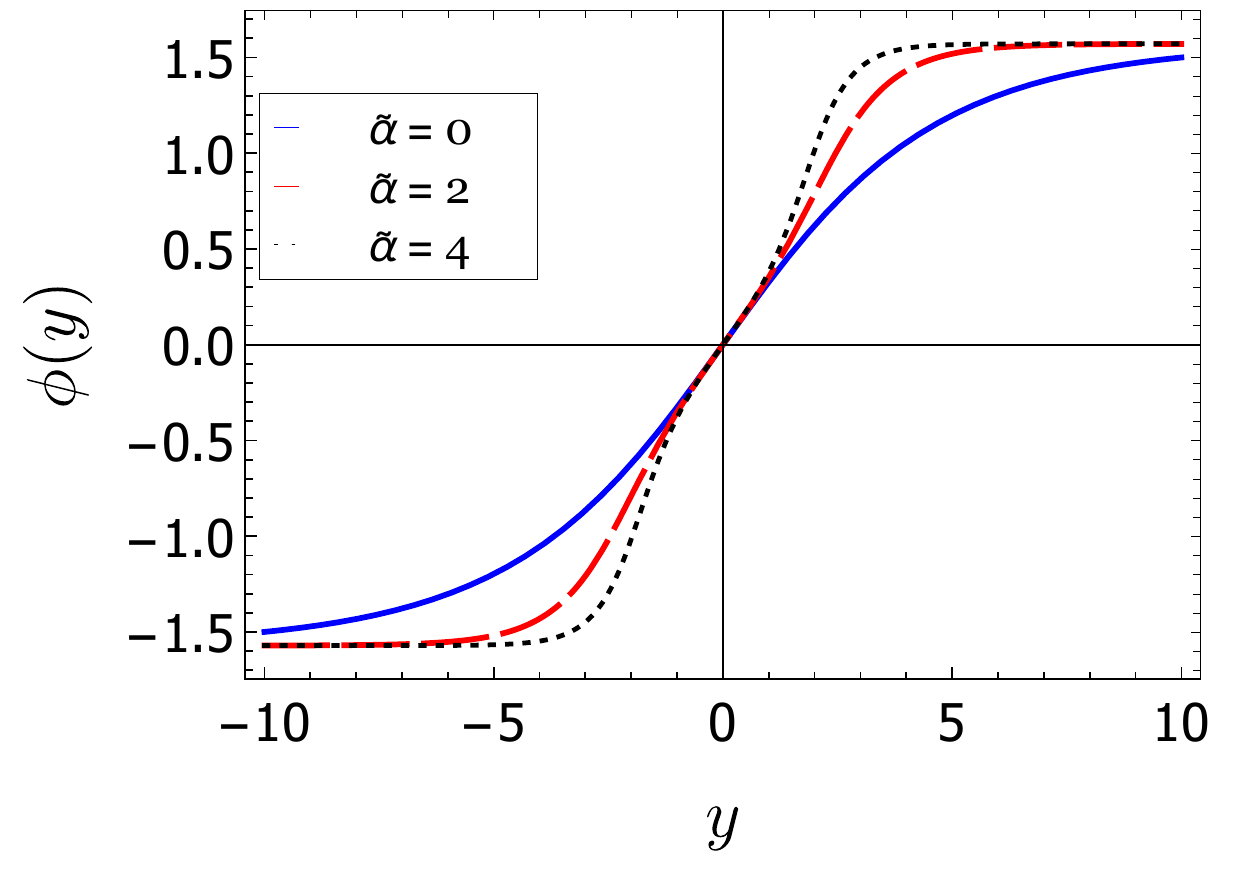}\qquad
\includegraphics[width=7cm,trim={0cm 0cm 0 0cm},clip]{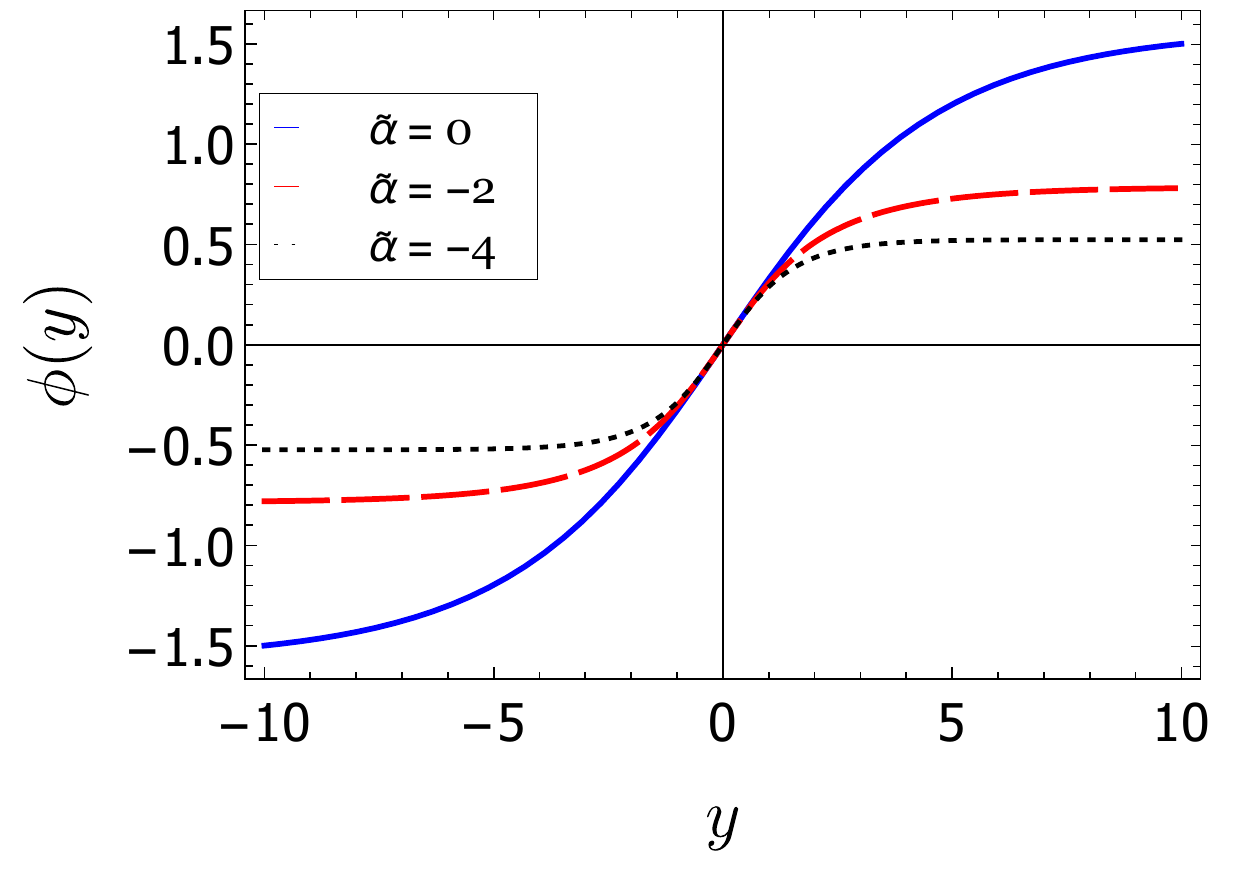}\\
\includegraphics[width=7cm,trim={0cm 0cm 0 0cm},clip]{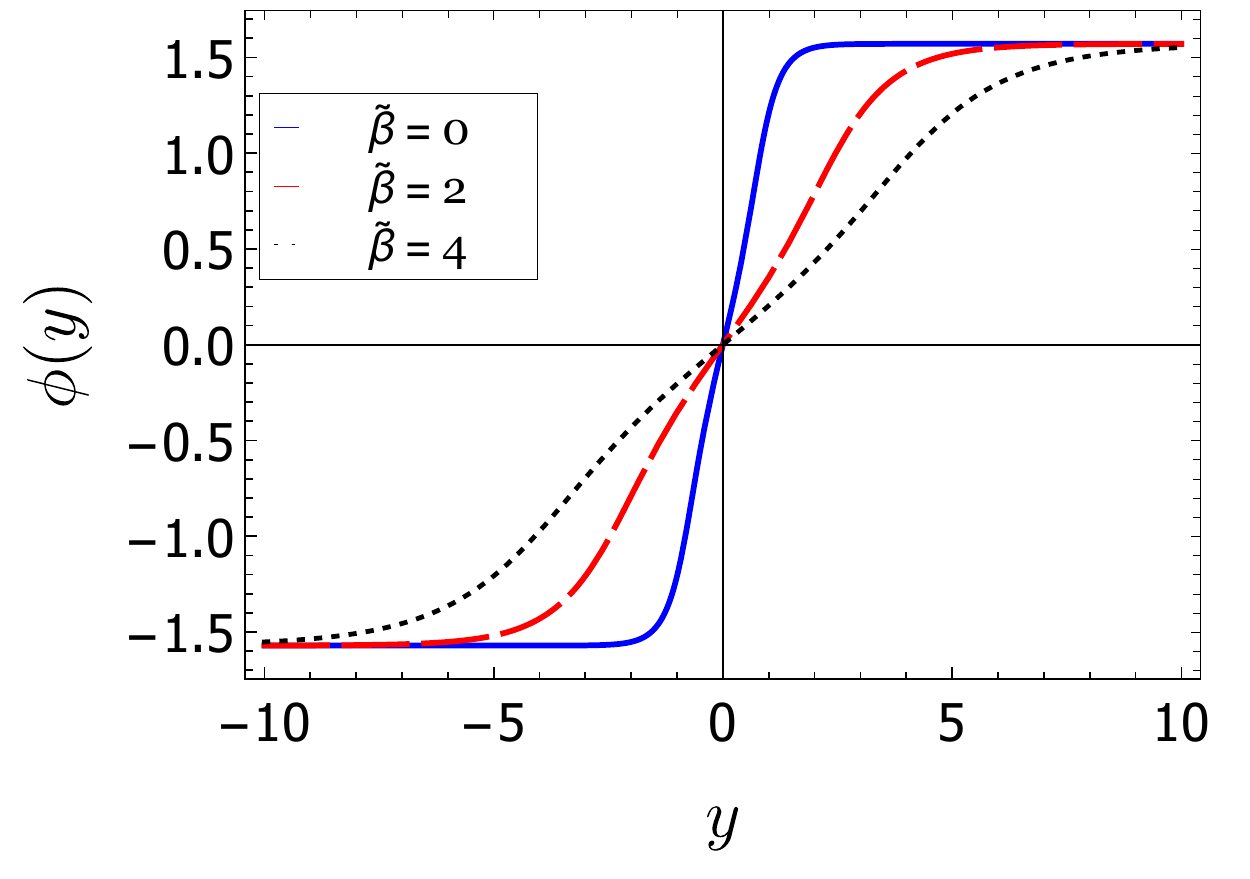}\qquad
\includegraphics[width=7cm,trim={0cm 0cm 0 0cm},clip]{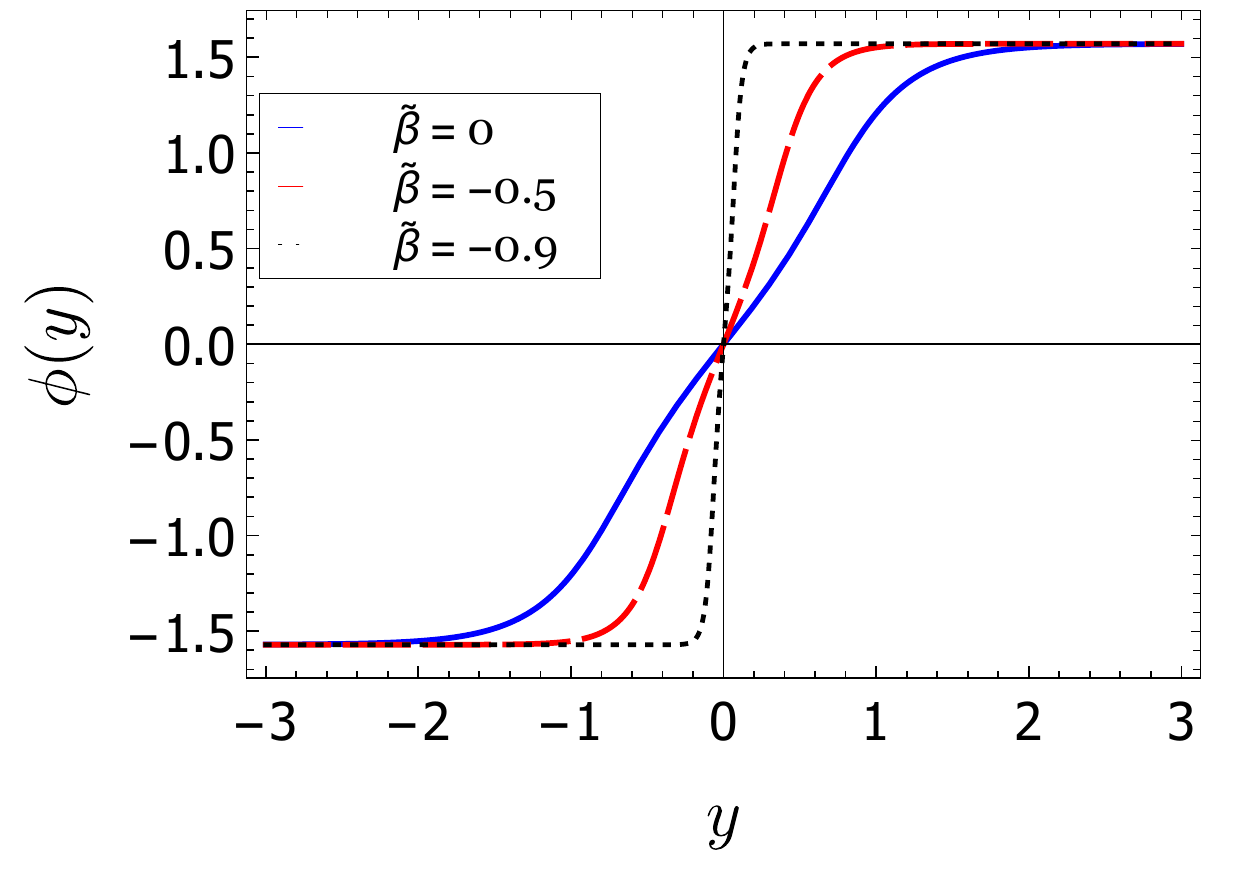}
\caption{Solution for the source field $\phi$ for $\tilde{\beta}=2$ and $\tilde{\alpha} = \{0, 2, 4\}$ (top left panel) or $\tilde{\alpha} = \{0, -2, -4\}$ (top right panel), and for $\tilde{\alpha}=2$ and $\tilde{\beta} = \{0, 2, 4\}$ (bottom left panel) or $\tilde{\beta} = \{0, -0.5, -0.9\}$ (bottom right panel).}
\label{fig1}
\end{figure*}

Before investigating the general case, let us consider first a specific case for which $f(R,T)\!=\!R+\beta T$ with $\beta$ a real parameter. In this case, the equation of motion for the source field simplifies to
\begin{eqnarray}\label{Eqmovphi}
\left(1\!+\!\frac{3\beta}{2 \kappa}\right)\left(\phi''+4A'\phi'\right)=\left(1\!+\!\frac{5\beta}{2\kappa}\right) \!V_\phi,
\end{eqnarray}
and the components of the modified Einstein field equations can be written as
\bes\label{EqEins2}
\bal
&\left(\!\kappa\!+\!\frac{3\beta}{2}\!\right)\!\phi'^2\!=\!-3 A''-12\alpha A'^2 A'' ,\label{EqEins2.1}\\
&\left(\!\kappa\!+\!\frac{5\beta}{2}\!\right)\! V\!=\!-\frac32A''\!-6A'^2\!-6\alpha A'^2\!\left(A''\!+\!2A'^2\right),\label{EqEins2.2}
\eal
\ees
where Eq. \eqref{EqEins2.2} is an adequate linear combination of Eqs. \eqref{EqEins2.1geral} and \eqref{EqEins2.2geral} used to eliminate the dependency of the result in $\phi'$. The equations that compose the system of Eqs. \eqref{Eqmovphi}, \eqref{EqEins2.1} and \eqref{EqEins2.2} are not all independent. To prove this statement, take the derivative of Eq. \eqref{EqEins2.2} with respect to $y$ and use Eq. \eqref{EqEins2.1} to get Eq. \eqref{Eqmovphi}. This implies that two of the previously mentioned equations are enough to fully describe the system, and one of the equations can be discarded without loss of information. As a result, one obtains a system of two independent equations for a total of three independent quantities, namely the source field $\phi$, the potential $V$, and the warp function $A$. This system is thus underdetermined and one extra constraint must be imposed to determine the system and allow one to obtain unique solutions. An interesting way to determine the system is through the first-order formalism, where we introduce an auxiliary function to reduce the order of the differential equations. Let us see how this works by introducing the first-order formalism through an auxiliary function $W(\phi)$ such that $A'=-\kappa W/3$. In such a case, the equations of motion \eqref{EqEins2} become
\bes
\bal
 \!\!\big(1+\tilde{\beta}\big)\phi'\!&=\!\left(1+\tilde{\alpha} W^2\right) W_\phi,\label{EqEins01}\\
\!\!\left(\!1\!+\!\frac{5\tilde{\beta}}{3}\right)V\!&=\!\frac{W_\phi^2\!\left(1\!+\!\tilde{\alpha}W^2\right)^2}{2 (1\!+\!\tilde{\beta})}\!-\!\frac{\kappa}{3} W^2\! \left(2\!+\!\tilde{\alpha} W^2\right),\label{EqEins02}
\eal
\ees
where we defined $W_\phi\equiv dW/d\phi$, $\tilde{\alpha}\equiv 4\alpha \kappa^2/9$ and $\tilde{\beta}\equiv 3\beta/(2\kappa)$. Note that this procedure has not yet determined the system, as we have added one more unknown, namely the function $W$. Solving the system of Eqs. \eqref{EqEins01} and \eqref{EqEins02} is, in general, a difficult task. Indeed, in most cases, one must recur to numerical methods in order to obtain solutions for the source field solution and the warp function. A interesting case in which analytical solutions are achievable is the sine-Gordon model, described by an auxiliary function in the form $W(\phi)=\sin(\phi)$. In this case, Eq. \eqref{EqEins01} yields
\begin{equation}
\big(1+\tilde{\beta}\big)\phi'=\big(1+\tilde{\alpha} \sin^2(\phi)\big) \cos(\phi) \,.
\end{equation}
Considering a boundary condition at the origin given by $\phi(0)=0$, the equation above yields a solution of the form
\begin{equation}\label{solphimod1}
\sqrt{\tilde{\alpha}}\arctan\!\left(\!\sqrt{\tilde{\alpha}}\sin(\phi)\right)\!+\!\ln\!\left(\!\tan\!\left(\frac{2\phi+\pi}{4}\right)\!\right)\!=\!\frac{1\!+\!\tilde{\alpha}}{1\!+\!\tilde{\beta}}\,y.
\end{equation}
Note that Eq. \eqref{solphimod1} provides a solution of $y$ as a function of $\phi$. To obtain the solution $\phi(y)$, this equation must be inverted, a procedure that can only be performed numerically. In Fig.~\ref{fig1}, we display the above solution for the source field for different values of $\tilde{\alpha}$ and $\tilde{\beta}$. We verify that for any positive value of $\alpha$ the asymptotic behavior of $\phi$ is $\phi_{\pm}\equiv\phi(y\to\pm\infty)\to\pm\pi/2$, but when $\alpha<0$ the value of $|\phi_\pm|$ decreases with an increase in $|\alpha|$. As for the parameter $\beta$, its value controls the slope of $\phi$, which becomes a step function in the limit $\beta\to-1$.

\begin{figure*}[t!]
\centering
\includegraphics[width=7cm,trim={0cm 0cm 0 0cm},clip]{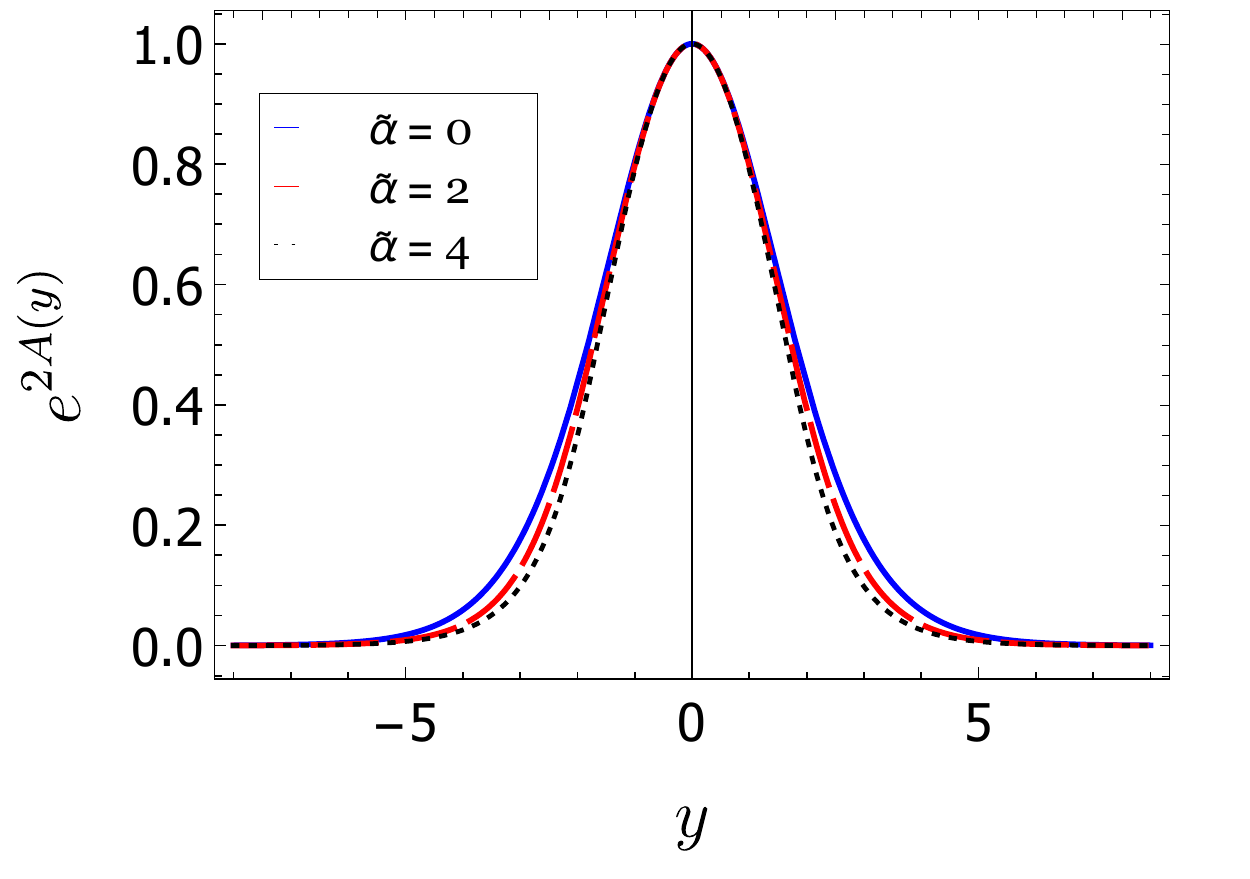}\qquad
\includegraphics[width=7cm,trim={0cm 0cm 0 0cm},clip]{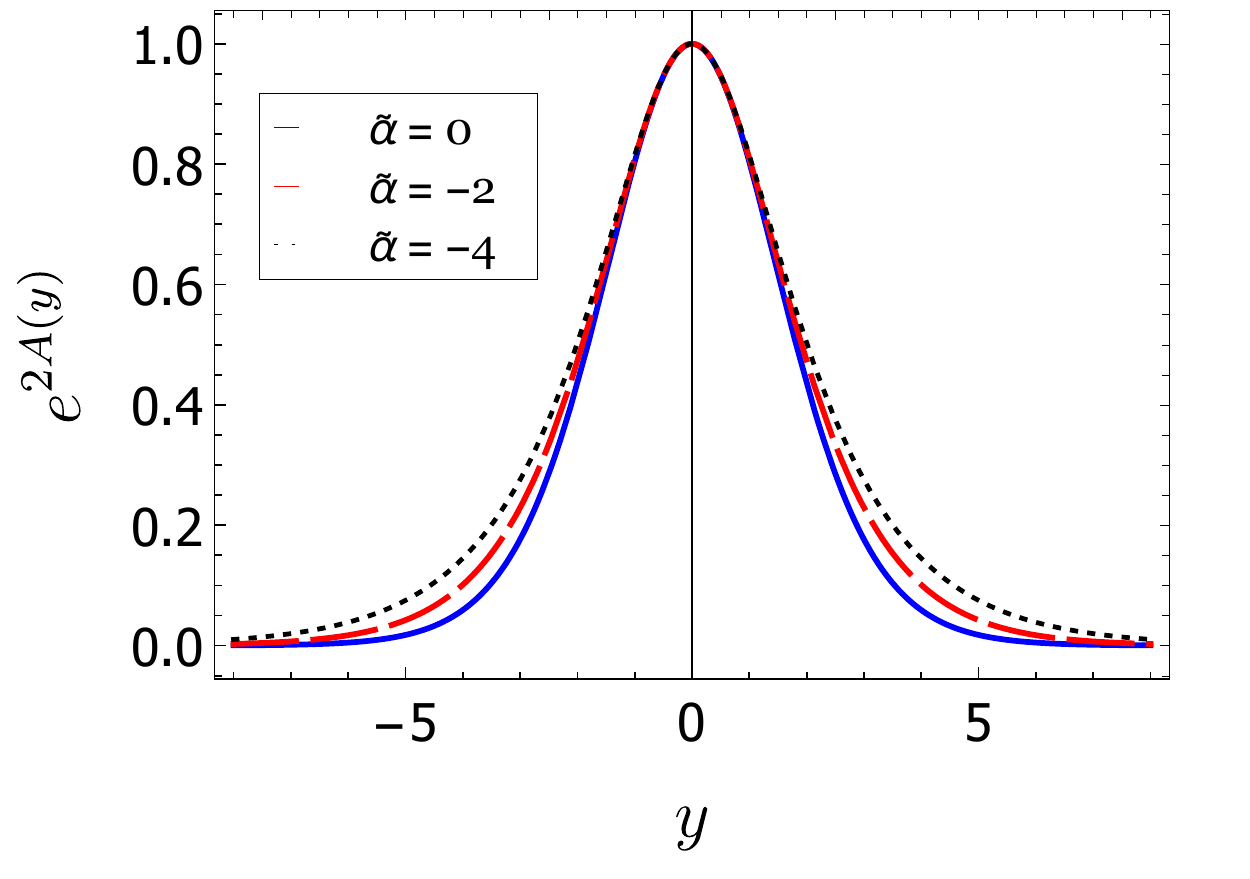}\\
\includegraphics[width=7cm,trim={0cm 0cm 0 0cm},clip]{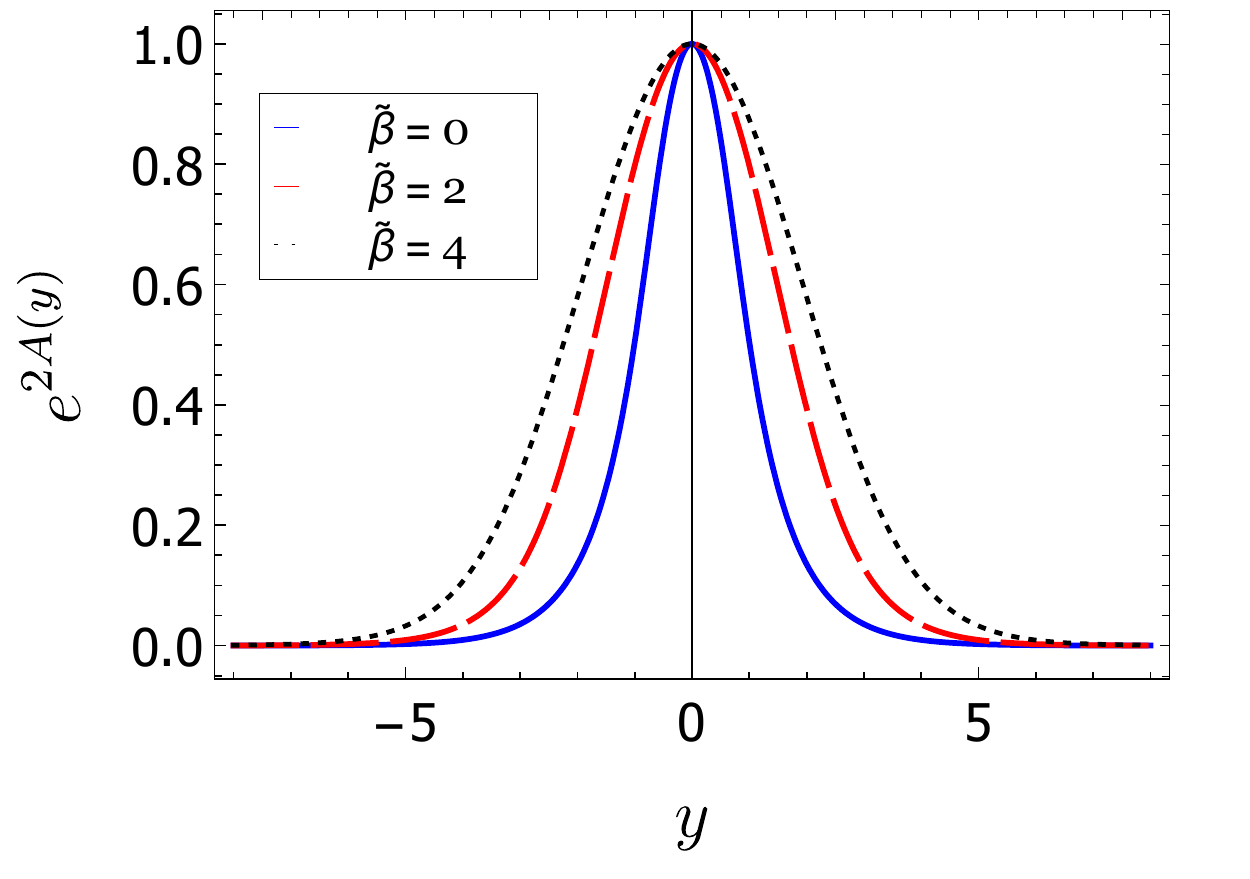}\qquad
\includegraphics[width=7cm,trim={0cm 0cm 0 0cm},clip]{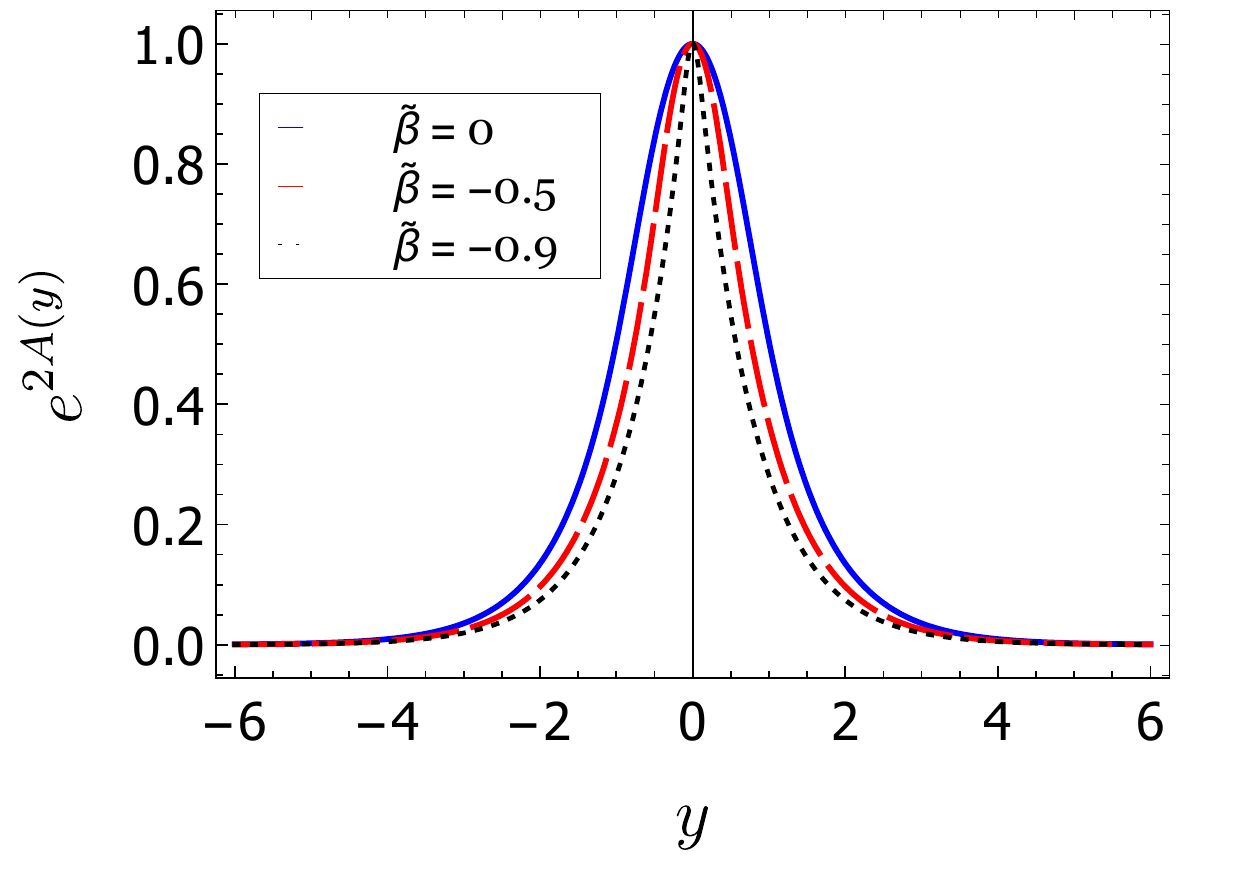}
\vspace{-0.3cm}
\caption{Warp factor $e^{2A}$ for $\tilde{\beta}=\kappa=2$ and $\tilde{\alpha} = \{0, 2, 4\}$ (top left panel) or $\tilde{\alpha} = \{0, -2, -4\}$ (top right panel), and for $\tilde{\alpha}=2$ and $\tilde{\beta} = \{0, 2, 4\}$ (bottom left panel) or $\tilde{\beta} = \{0, -0.5, -0.9\}$ (bottom right panel).}
\label{fig2}
\end{figure*}

The warp function can then be written as
\begin{equation}\label{solwarpmod1}
    A(\phi)=-\frac{\kappa}{6(1+\tilde{\alpha})}\ln\left(\frac{2+\tilde{\alpha}-\tilde{\alpha} \cos(2\phi)}{2\cos^2(\phi)}\right),
\end{equation}
where we have imposed the boundary condition $A(0)=0$. The origin and asymptotic values of the Kretschmann scalar are given by $K(0)=16\kappa^2/9$ and $K(\phi_{\pm})=40\kappa^4/81$, respectively. In Fig.~\ref{fig2}, we display warp factor $e^{2A}$ for $\kappa=2$ and for the same values of $\tilde{\alpha}$ and $\tilde{\beta}$ used in Fig.~\ref{fig1}. We verify that the parameters $\alpha$ and $\beta$ have a similar but opposite effect on the warp function: an increase in $\alpha$ narrows the warp function, while an increase in $\beta$ widens it. Furthermore, the parameter $\alpha$ changes the width of the warp function far from $y=0$ while keeping the behavior at $y=0$ virtually unchanged, while the parameter $\beta$ affects the whole range of $y$ at once.

The potential of the source field of the brane can be written as
\be
\begin{aligned}
    V(\phi)=\,&\frac{3}{5\tilde{\beta}+3} \Bigg(\frac{\cos^2(\phi) \big(1+\tilde{\alpha}\sin ^2(\phi)\big)^2}{2 (\tilde{\beta}+1)}\\
    &-\frac{2}{3} \sin^2(\phi) \big(2+\tilde{\alpha}\sin ^2(\phi)\big)\Bigg).
\end{aligned}
\ee
This potential can be associated to a five-dimensional cosmological constant, whose explicit form can be obtained by $\Lambda_5\equiv V(\phi_{\pm})=-2(2+\tilde{\alpha})/(3+5\tilde{\beta})$. For $\tilde\alpha>-2$ and $\tilde\beta>-3/5$, or $\tilde\alpha<-2$ and $\tilde\beta<-3/5$, we have $\Lambda_5<0$, implying that the bulk is asymptotically $AdS_5$, whereas for $\tilde\alpha>-2$ and $\tilde\beta<-3/5$, or $\tilde\alpha<-2$ and $\tilde\beta>-3/5$, we have $\Lambda_5>0$, implying that the bulk is asymptotically $dS_5$. If $\tilde{\alpha}=-2$ we have $\Lambda_5=0$ and the bulk is asymptotically Minkowski. 

\section{Scalar-tensor representation}\label{sec_sca_tens}

Let us now investigate the general case of the $f(R,T)-$brane. An interesting way to perform this analysis is through the scalar-tensor representation, where the arbitrary dependence of the action in $R$ and $T$ is exchanged by two auxiliary scalar fields. We start by introducing two auxiliary scalar fields $\psi$ and $\chi$ such that 
\begin{equation}\label{defscalar}
\psi=\frac{\partial f}{\partial R}, \qquad\chi=\frac{\partial f}{\partial T}.
\end{equation}
Here, the auxiliary fields are coupled via a potential $U(\psi,\chi)$ as
\begin{equation}
U\left(\psi,\chi\right)=\psi R+\chi T-f\left(R,T\right).
\end{equation}
In the scalar-tensor representation, the auxiliary fields $\psi$ and $\chi$ carry the information about the dependency of the function $f\left(R,T\right)$ in $R$ and $T$. As show in \cite{Rosa:2021tei}, the introduction of these auxiliary fields allows one to rewrite the action \eqref{actiongeo} into a dynamically equivalent scalar-tensor representation in the form 
\begin{eqnarray}
S\!=\!\frac{1}{2\kappa}\!\int_\Omega\!\!\!\sqrt{|g|}\left(\psi R\!+\!\chi T\!-\!U\!+\!\alpha G\right)d^5x\!+\!S_m\!\left(\phi,g_{ab}\right).\label{actionst}
\end{eqnarray}
Now, each field represents an additional degree of freedom, leading to a larger number of equations of motion. In this new representation, the modified Einstein field equations take the form
\be\label{EinsEqST}
\begin{aligned}
&\psi R_{ab}+\left(g_{ab}\Box -\nabla_a\nabla_b\right)\psi-\frac12g_{ab}\left(\psi R-U\right)\\
&+\alpha H_{ab}=\kappa T_{ab}-\left(T_{ab}+\Theta_{ab}+\frac12g_{ab}T\right)\chi,
\end{aligned}
\ee
where $H_{ab}$ was previously defined in Eq. \eqref{hab}. Taking a variation of Eq. \eqref{actionst} with respect to the scalar fields $\psi$ and $\chi$ yields respectively $U_\psi=R$ and $U_\chi=T$. For static configurations, such that $\psi=\psi(y)$ and $\chi=\chi(y)$, we get $U_\psi=20A'^2+8A''$ and $U_\chi=(3/2)\phi'^2+5V$. Using the chain rule on the auxiliary potential as $U'=U_\psi\psi'+U_\chi\chi'$ we find
\begin{equation}\label{Urelfinal}
U'=\left(20A'^2+8A''\right)\psi'+\left(\frac32\phi'^2+5V\right)\chi'.
\end{equation}
On the other hand, the independent components of the modified Einstein's field equations become
\bes\label{EqEins2.ST}
\bal
&-3A''\psi+A'\psi'-\psi''-12\alpha A'^2 A''\!=\!\left(\!\kappa\!+\!\frac{3\chi}{2}\!\right)\phi'^2,\label{EqEins2.1ST}\\
&-3A'^2\psi+\frac14U-2A'\psi'-6\alpha A'^4\nn
&=-\frac14\left(\kappa+\frac{3\chi}{2}\right)\phi'^2\!+\!\frac12\left(\kappa+\frac{5\chi}{2}\right)V.\label{EqEins2.2ST}
\eal
\ees
Moreover, in this representation the equation of motion for the source field $\phi$ takes the form
\begin{eqnarray}\label{EqmovphiET}
\left(\!1\!+\!\frac{3\chi}{2 \kappa}\!\right)\left(\phi''\!+\!4A'\phi'\right)\!+\!\frac{3\chi'}{2\kappa }\phi'\!=\!\left(\!1\!+\!\frac{5\chi}{2\kappa}\!\right)\! V_\phi.
\end{eqnarray}
It is important to note that only three of the four equations above are independent and thus one of them can be discarded without loss of information. For the sake of simplicity, we chose to discard Eq. \eqref{EqEins2.2ST} and keep Eqs. \eqref{Urelfinal}, \eqref{EqEins2.1ST} and \eqref{EqmovphiET} as the set of three independent equations to be solved. On the other hand, the number of unknowns in the system is six, namely $A(y)$, $\phi(y)$, $V$, $\psi(y)$, $\chi(y)$ and $U$. It is evident that the number of variables exceeds the number of equations by three. This allows us to impose three additional restrictions to determine the system. Here, we follow the strategy used in \cite{Rosa:2022fhl,Bazeia:2022agk} and assume that the brane background is identical to the standard case, having the source field $\phi$, the warp function $A$ and the potential $V$ the same solutions as obtained in the absence of the fields of the scalar-tensor representation and the Gauss-Bonnet term, i.e. with $\alpha=0$ and $f(R,T)=R$. This choice ensures the asymptotic behavior of the solutions and facilitates the resolution of the equations of motion of the scalar fields. Furthermore, these assumptions preserve the regularity of the Ricci scalar and the Kretschmann scalar.

In this case, we can also investigate brane models using the first-order formalism. For this purpose, we introduce a function $W$ such that
\be\label{FOF}
    \phi'=W_\phi,\quad A^{\prime}=-\frac{\kappa}3W(\phi),\quad V=\frac12W_\phi^2-\frac{\kappa}3W^2.
\ee
These assumptions reduce the number of degrees of freedom by two, as they introduce one extra quantity $W$ and three extra constraints for $\phi$, $A$ and $V$. Inserting the definitions in Eq. \eqref{FOF} into Eq.~\eqref{EqmovphiET}, we obtain
\begin{eqnarray}\label{eqpsi}
9\chi'+2\left(4\kappa W-3W_{\phi\phi}\right)\chi=0.
\end{eqnarray}
With this, the solution for the field $\chi$ can be obtained as
\begin{eqnarray}\label{eqpsisol}
\chi(y)=\chi_0\exp\left(-\frac29\int dy\left(4\kappa W-3W_{\phi\phi}\right)\right),
\end{eqnarray}
where $\chi_0=\chi(0)$ is an integration constant. Inserting the first-order equations into Eqs. \eqref {EqEins2.1ST} and \eqref {Urelfinal} we obtain
\begin{equation}\label{eqFOst1}
\psi ''\!+\!\frac{\kappa}{3}W\psi'\!-\!\kappa W_{\phi}^2\psi\!=\!\kappa W_{\phi}^2 \!\left(\!\frac{4\alpha\kappa^2}9 W^2\!-\!\frac{3\chi}{2\kappa}\!-\!1\!\right),
\end{equation}
\begin{equation}\label{eqFOst2}
    U'=\frac{4\kappa}{9}\left(5\kappa W^2-6 W_{\phi}^2\right)\psi'-\frac{2}{3} \left(5\kappa W^2-6 W_{\phi}^2\right) \chi' \,.
\end{equation}
At this point we must properly choose the function $W(\phi)$ to determine the system and  solve the equations of motion. Given the degree of complexity of the system under consideration, no analytical solutions can be found and we must recur to numerical methods in what follows. As a next step, let us investigate some specific models.

\section{Specific Models}\label{models}

\subsection{Model A}
As first model, let us consider $W=\phi-(1/3)\phi^3$. Using the first-order formalism we obtain the solution as $\phi(y)=\tanh(y)$. Moreover, the warp function becomes
\begin{equation}
    A(y)=  \frac{2\kappa}{9} \ln (\sech(y))-\frac{\kappa}{18}\tanh^2(y).
\end{equation}
On the other hand, the potential $V$ takes the form
\begin{equation}
    V(\phi)=\frac{1}{2} \left(\phi^2-1\right)^2-\frac{\kappa}{27}\phi^2\left(\phi^2-3\right)^2.
\end{equation}
As the background equations are the same as in the standard case, that is, with $\alpha=0$ and $f(R,T)=R$, we do not need to investigate the behavior of the Kretschmann scalar. Moreover, the five-dimensional cosmological constant leads to a bulk which is always asymptotically $AdS_5$.

\begin{figure}[t!]
\centering
\includegraphics[width=7cm,trim={0cm 0cm 0 0cm},clip]{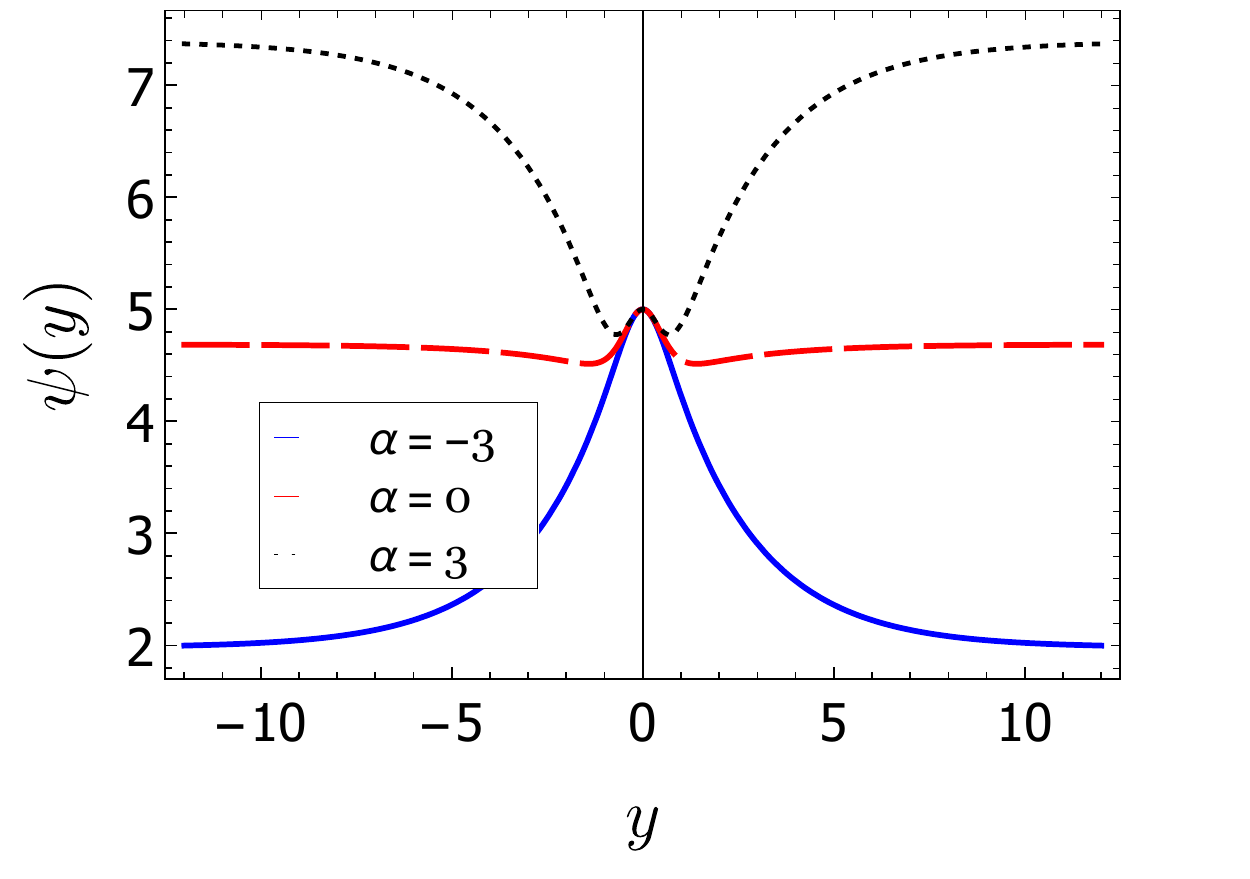}
\includegraphics[width=7cm,trim={0cm 0cm 0 0cm},clip]{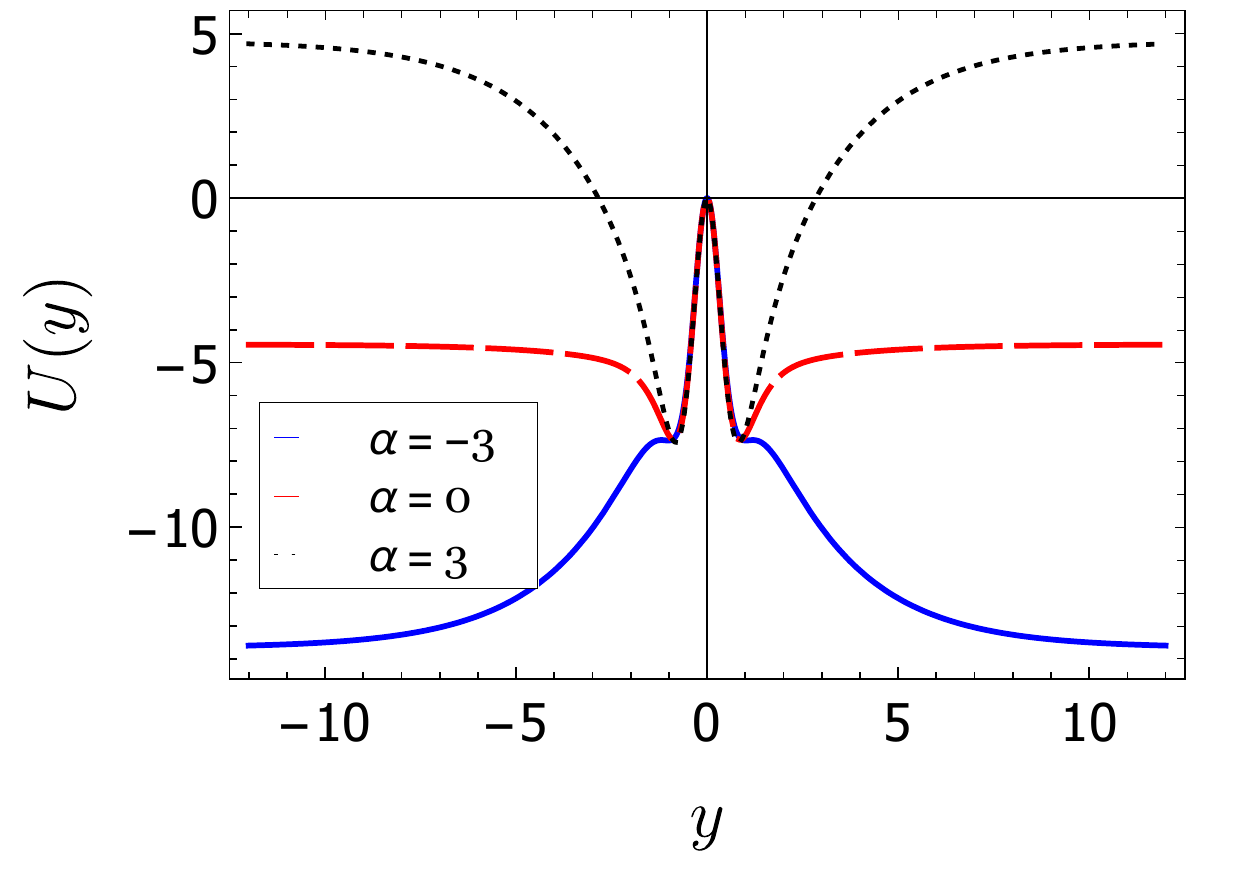}
\vspace{-0.3cm}
\caption{Scalar field $\psi$ in model A for $\alpha =\{-3, 0, 3\}$ (top panel). Auxiliary potential $U(y)$ for the same values of $\alpha$ (bottom panel). In both cases we use $\kappa=2$, $\psi(0)=5$, $\chi(0)=7$, and $U(0)=0$.}
\label{fig3}
\end{figure}

The solution for the scalar field $\chi$ can be obtained by taking the integral of Eq. \eqref{eqpsisol} which yields
\begin{eqnarray}
\chi(y)=\chi _0 \frac{e^{-(4\kappa/27) \tanh ^2(y)}}{\cosh ^{4(4 \kappa +9)/27}(y) }.
\end{eqnarray}
This implies that $\chi$ is not affected by the parameter $\alpha$ and the result is identical to that obtained in \cite{Bazeia:2022agk}. However, the $\alpha$ parameter is present in the equation for $\psi$. With this, we can perform the numerical integration of Eqs. \eqref{eqFOst1} and \eqref{eqFOst2}. The boundary conditions at $y=0$, especially for $\psi(0)$ and $\chi(0)$, strongly modify the behavior of the auxiliary fields. However, since the effect of a variation in these boundary conditions was already previously analyzed in Ref. \cite{Rosa:2021tei}, in this work we keep these boundary conditions constant and analyze solely the effects of a variation in $\alpha$, which also plays an important role in the qualitative behavior of the solutions. In Fig.~\ref{fig3}, we display the field $\psi$ and the auxiliary potential $U$ for different values of the parameter $\alpha$, while keeping the boundary conditions at $\psi(0)=5$, $\chi(0)=7$, and $U(0)=0$. We observe that variations in the value of $\alpha$ affect the asymptotic value of the solutions without altering their behavior in the vicinity of $y=0$. Consequently, the potential $U$ might feature different combinations of potential wells and barriers depending on the value of $\alpha$ and on the background choice of boundary conditions, which then manifests as different distributions of the field $\chi$.

\subsection{Model B}
\begin{figure}[t!]
\centering
\includegraphics[width=7cm,trim={0cm 0cm 0 0cm},clip]{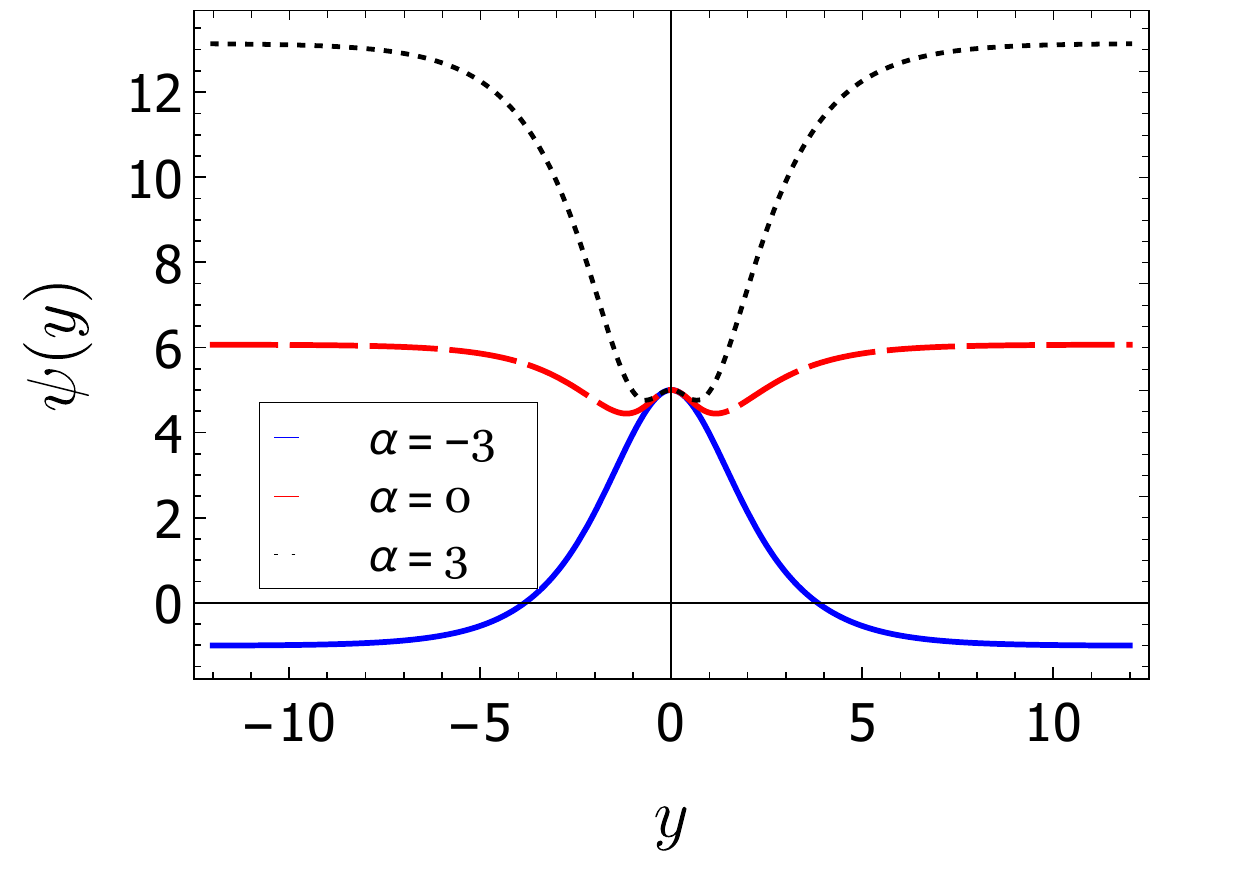}
\includegraphics[width=7cm,trim={0cm 0cm 0 0cm},clip]{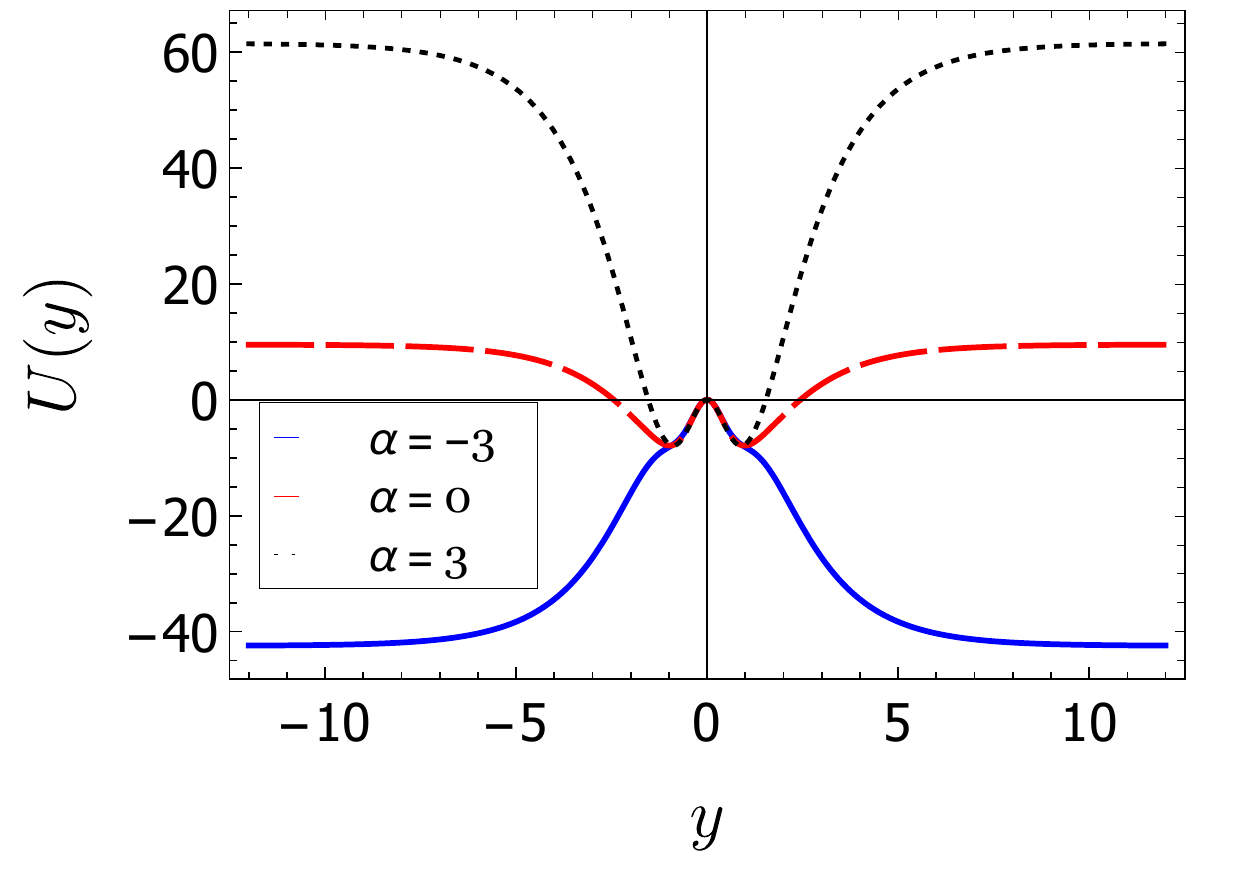}
\vspace{-0.3cm}
\caption{Scalar field $\psi$ in model B for $\alpha =\{-3, 0, 3\}$ (top panel). Auxiliary potential $U(y)$ for the same values of $\alpha$ (bottom panel). In both cases we use $\kappa=2$, $\psi(0)=5$, $\chi(0)=7$, and $U(0)=0$.}
\label{fig4}
\end{figure}

Let us now consider again the sine-Gordon Model for which $W=\sin(\phi)$. By using Eqs. \eqref{FOF} we obtain the solution for the source field as $\phi(y)=2\arctan(\tanh(y/2))$. Moreover, the warp function becomes
\begin{equation}
    A(y)=  \frac{\kappa}{3} \ln \left(\sech(y)\right).
\end{equation}
On the other hand, the expression for the potential $V$ is given by
\begin{equation}
    V(\phi)=\frac{1}{12} ((2 \kappa +3) \cos (2 \phi )-2 \kappa +3).
\end{equation}
Once again, we emphasize that these results are known and lead to a regular Kretschmann scalar and for the correct asymptotic behavior of the brane. Therefore, from Eq. \eqref{eqpsisol} we obtain 
\begin{eqnarray}
\chi(y)=\chi_0\, \sech^{2(4 \kappa +3)/9}(y),
\end{eqnarray}
where again we use the boundary condition at the origin given by $\chi(0)=\chi_0$. Plugging all these results into Eqs. \eqref{eqFOst1} and \eqref{eqFOst2} we can solve the system to obtain the solutions for the auxiliary field $\psi$ and potential $U$. In Fig.~\ref{fig4}, we display the field $\psi$ and the auxiliary potential $U(y)$ for the sine-Gordon model, for the same boundary conditions as chosen in the previous model, i.e., $\psi(0)=5$, $\chi(0)=7$, and $U(0)=0$. These results follow closely the behaviors presented for model A, i.e., the value of $\alpha$ alters the asymptotic value of the solutions without altering the behavior in the vicinity of $y=0$, thus allowing one to induce qualitatively different solutions controlled by the parameter $\alpha$.

\subsection{Model C}

\begin{figure}[t!]
\centering
\includegraphics[width=7cm,trim={0cm 0cm 0 0cm},clip]{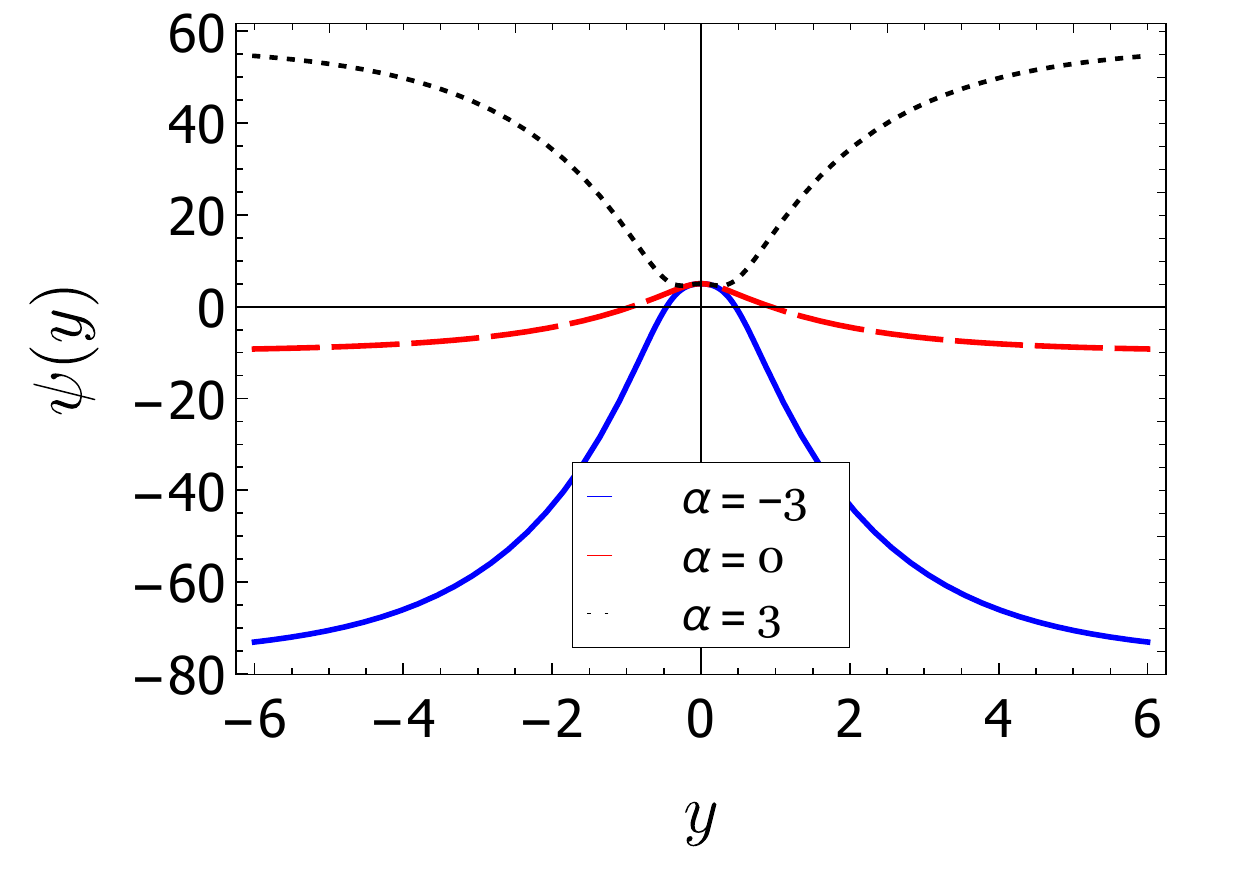}
\includegraphics[width=7cm,trim={0cm 0cm 0 0cm},clip]{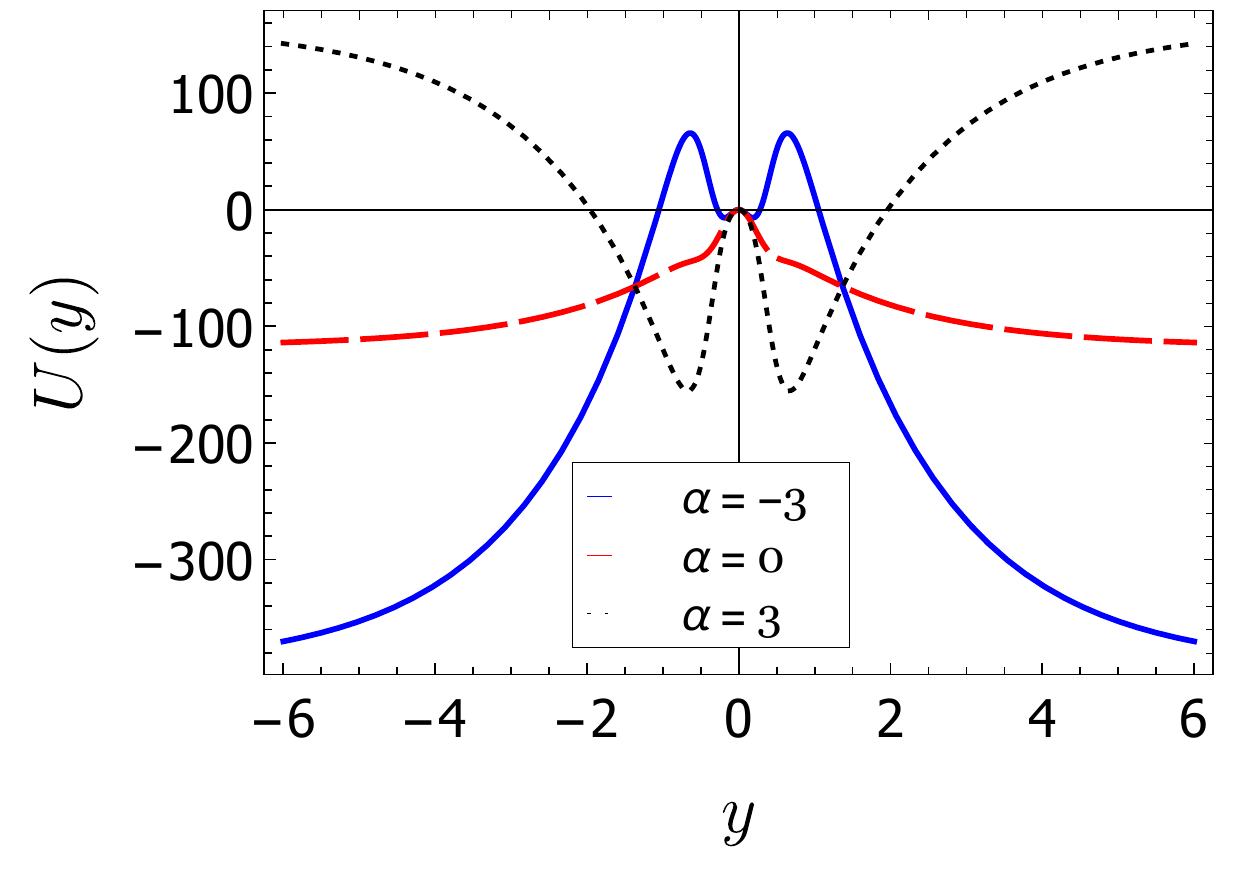}
\vspace{-0.3cm}
\caption{Scalar field $\psi$ in model C for $\alpha =\{-3, 0, 3\}$ (top panel). Auxiliary potential $U(y)$ for the same values of $\alpha$ (bottom panel). In both cases we use $\kappa=2$, $\psi(0)=5$, $\chi(0)=7$, and $U(0)=0$.}
\label{fig5}
\end{figure}

As can be verified from Eqs. \eqref{EqEins2.ST}, the Gauss-Bonnet term introduces additional contributions that depend on derivatives of the warp function. Therefore, it is natural to assume that a $W(\phi)$ that induces a strong modification in the warp function can lead to a behavior different from those obtained in previous models, thus causing a greater impact on the behavior of the auxiliary fields. A model with these characteristics was investigated in \cite{Bazeia:2015eta}, where the authors proposed a $W(\phi)$ as
\begin{equation}\label{superMC}
    W(\phi)=-\frac1{\sqrt{\lambda}(1-\lambda)}\ln\left(\frac{1-\sqrt{\lambda}\,\sn(\phi,\lambda)}{\dn(\phi,\lambda)}\right),
\end{equation}
where $\lambda$ is a real parameter in the range $[0,1)$ and $\sn$ and $\dn$ are the Jacobi's elliptic functions. Using the first-order equations, we get the source field solution as
\begin{equation}\label{eqsolModC}
    \phi(y)=\sn^{-1}\left(\tanh\left(\frac{y}{1-\lambda}\right),\,\lambda\,\right).
\end{equation}
In this case, the warp function is obtained numerically. The solutions for the field $\chi$ under variations of the parameter $\lambda$ in this model have already been studied and analyzed previously in Ref. \cite{Bazeia:2022agk}, and thus we choose not to repeat that analysis here. Thus, in this work we retain the value of $\lambda$ constant throughout the analysis and study solely the effects of the parameter $\alpha$. In Fig. \ref{fig5} we represent the field $\psi$ and the potential $U(y)$ for the same boundary conditions chosen in the previous models, i.e., $\psi(0)=5$, $\chi(0)=7$, and $U(0)=0$. Again, we verify that the effects of the parameter $\alpha$ are similar to the previous models, i.e., it affects the asymptotic values of the solutions while keeping the behavior close to $y=0$ unaltered. The main difference between this model and the previous ones is that the behavior of the solutions in a vicinity of $y=0$ presents a simple structure.

\section{Linear Stability}\label{stabiline}

In this section, we investigate the linear stability of the $f(R,T)$-gravity in the presence of a Gauss-Bonnet term. We start by considering small perturbations around the metric background in the form
\be
ds^2=e^{2A}(\eta_{\mu\nu}+h_{\mu\nu}(x^\sigma,y))dx^{\mu}dx^{\nu}-dy^2,
\ee
where any tensor component in the extra dimension is taken to be zero and $x^\sigma$ is a vector in the four-dimensional space. As usual, we consider the transverse-traceless (TT) gauge condition for the four-dimensional component $h_{\mu\nu}$, i.e., $\partial_\mu h^\mu_{\nu}=0$ and $h^\mu_{\mu}=0$, and we raise or lower indices as $h_{\mu\nu}=-\eta_{\mu\alpha}\eta_{\nu\beta}h^{\alpha\beta}$. Considering only first-order perturbations, the modified Einstein field equations \eqref{EinsEq} can be written as
\be
\begin{aligned}
&(f_{RR}\delta R+f_{RT}\delta T)R_{ab}+f_{R}\delta R_{ab}+\delta g_{ab}\Box f_R\\
&\!+g_{ab}\delta(\Box f_R)\!-\!\delta(\nabla_a\nabla_bf_{R})\!-\!\frac12g_{ab}(f_{R}\delta R\!+\!f_{T}\delta T)\\
&-\frac12\delta g_{ab}f+\alpha \delta H_{ab}=\kappa \delta T_{ab}-\big(\delta T_{ab}+\delta \Theta_{ab}\big)f_T\\
&-\big(T_{ab}+\Theta_{ab}\big)(f_{RT}\delta R+f_{TT}\delta T).\label{comppertub}
\end{aligned}
\ee

In general, the linear perturbation can be decomposed into scalar, vector and tensor parts, which are represented by the components $44$, $\mu 4$ and $\mu\nu$ in the equation above. However, if we use the transverse-traceless conditions, only the tensor perturbations remain. 

To deal with perturbations in the Riemann tensor we can use the relation $\delta (R_{abc}{}^d)=\nabla_b\delta\Gamma_{ac}^d-\nabla_a\delta\Gamma_{bc}^d$. Furthermore, we can construct the non-zero perturbations in the Christoffel symbols as
\bes
\bal
&\delta\Gamma_{\mu\nu}^5=\frac12\big(e^{2A}h_{\mu\nu}\big)',\\
&\delta\Gamma_{\mu\nu}^\sigma=\frac12\eta^{\sigma\lambda}\big(\partial_\nu h_{\mu \lambda}+\partial_\mu h_{\nu\lambda}-\partial_\lambda h_{\mu\nu}\big),\\
&\delta\Gamma_{\mu 5}^\sigma=\frac12\eta^{\sigma\lambda}h_{\mu\lambda}'.
\eal
\ees
We thus obtain that $\delta R=\delta(\Box f_R)=0$. Furthermore, $\delta(\nabla_\mu\nabla_\nu f_{R})=-\delta\Gamma_{\mu\nu}^5f_R'$ and
\be
\begin{aligned}
\delta R_{\mu\nu}=&\,e^{2A}\!\left(\frac12h_{\mu\nu}''+2A'h_{\mu\nu}'+(A''+4A'^2)h_{\mu\nu}\right)\\
&-\frac12\Box^{(4)} h_{\mu\nu},
\end{aligned}
\ee
\be
\begin{aligned}
\delta H_{\mu\nu}\!=&-\!12e^{2A}\!A'^2 \!\left(\!A''\!\!+\!A'^2\right)\!h_{\mu\nu}\!-\!2\!\left(\!A''\!\!+\!A'^2\right) \!\Box^{(4)} h_{\mu\nu}\\
&+2e^{2A}A'^2 h_{\mu\nu}''\!+\!4 e^{2A} A' \!\left(\!A''\!\!+\!2 A'^2\right)\!h_{\mu\nu}',
\end{aligned}
\ee
where $\Box^{(4)}=\eta^{\alpha\beta}\partial_\alpha\partial_\beta$. On the other hand, we can follow the prescription adopted in \cite{Csaki:2000fc} and consider only perturbations in $h_{\mu\nu}$ in the tensors $\delta T_{\mu\nu}$ and $\delta \Theta_{\mu\nu}$, and we can write,
\bes
\bal
&\delta T_{\mu\nu}=\frac14\eta^{\alpha\beta}T_{\alpha\beta}h_{\mu\nu},\\
&\delta \Theta_{\mu\nu}=\frac14\eta^{\alpha\beta}\Theta_{\alpha\beta}h_{\mu\nu}.
\eal
\ees
Consequently, we must make $\delta T=0$ in Eq. \eqref{comppertub}. With this, we can write the four-dimensional component of Eq. \eqref{comppertub} as
\be
\begin{aligned}
a(y)h_{\mu\nu}''+e^{-4A}\big(e^{4A}a(y)\big)'h_{\mu\nu}'=e^{-2A}b(y)\Box^{(4)} h_{\mu\nu},
\end{aligned}
\ee
where $a(y)=f_{R}+4\alpha A'^2$ and $b(y)=f_{R}+4\alpha (A'^2+A'')$. At this point, we can introduce a coordinate transformation as $dz=e^{-A}dy$ and consider the decomposition in $h_{\mu\nu}$ as $h_{\mu\nu}(x^\sigma,z)=e^{-3A/2}a^{-1/2}H_{\mu\nu}(z)P(x^\sigma)$, where $P(x^\sigma)$ satisfies the Klein-Gordon equation, i.e, $\Box^{(4)} P(x^\sigma)=-\omega^2 P(x^\sigma)$. We thus obtain a Schrodinger-like equation as
\be\label{ShorinEq}
-\frac{d^2}{dz^2}H_{\mu\nu}+{\cal U}(z)H_{\mu\nu} =\frac{b(z)}{a(z)}\omega^2 H_{\mu\nu},
\ee
where
\be
{\cal U}(z) =\frac12\frac{3A_za_z+a_{zz}}{a}+\frac34\left(2A_{zz}+3A_z^2\right)-\frac{a_z^2}{4a^{2}}.
\ee
Equation \eqref{ShorinEq} can be written as
\be
{\cal S}^{\dag}{\cal S}H_{\mu\nu} =\frac{b(z)}{a(z)}\omega^2 H_{\mu\nu},
\ee
where ${\cal S}=-\partial_z+a_z/(2a)+3A_z/2$. Note that if $b(z)/a(z)\geq0$ the system is stable under tensor perturbations. In particular, the zero mode, obtained from $\omega=0$ has the form
\be
H_{\mu\nu}^{(0)}=N_{\mu\nu} \sqrt{a(z)}\, e^{3A/2},
\ee
where $N_{\mu\nu}$ is the normalization constant. Using the scalar-tensor representation, we can represent the zero mode as
\be
H_{\mu\nu}^{(0)}(z)=N_{\mu\nu} \, e^{3A/2}\sqrt{\psi(z)+4\alpha A_z^2e^{-2A}}.
\ee

\begin{figure*}[t!]
\centering
\includegraphics[width=7cm,trim={0cm 0cm 0 0cm},clip]{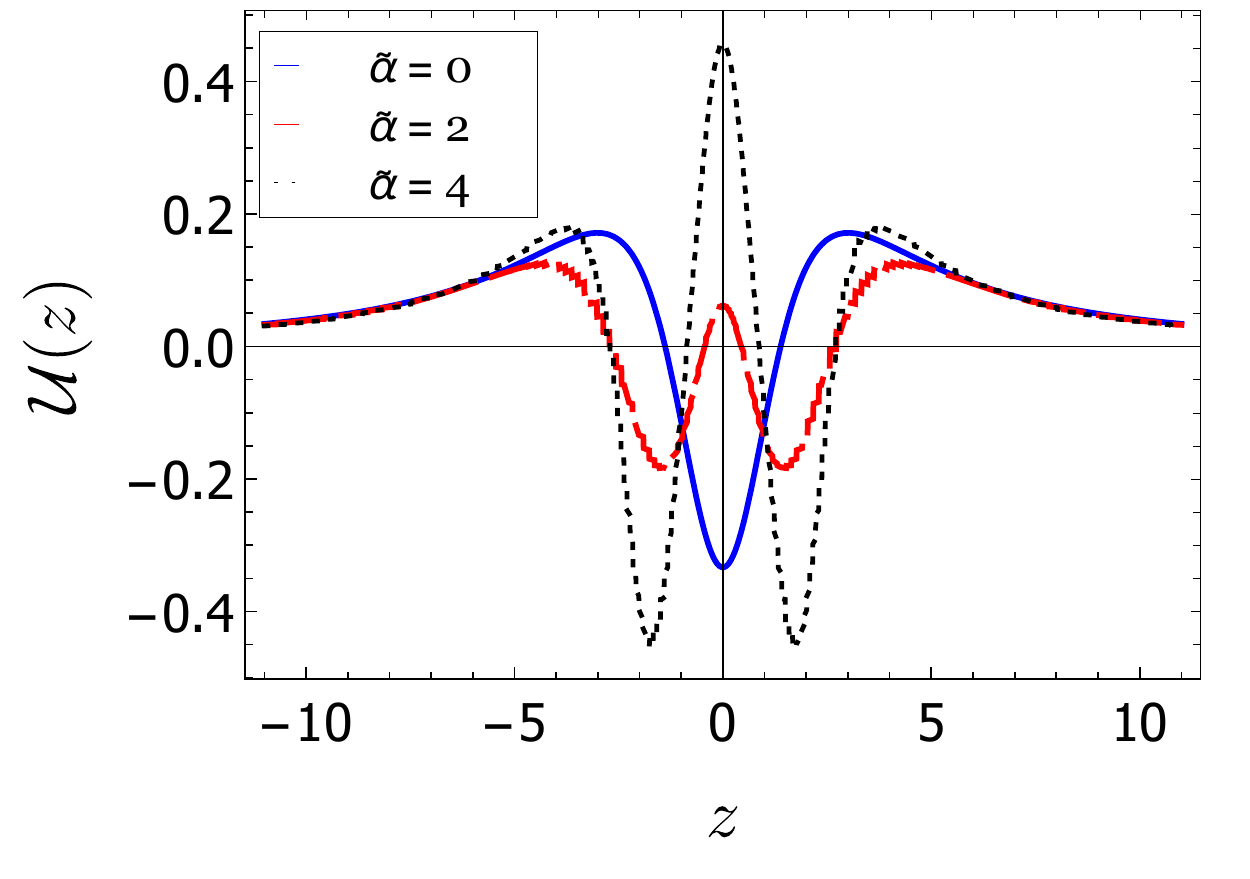}
\includegraphics[width=7cm,trim={0cm 0cm 0 0cm},clip]{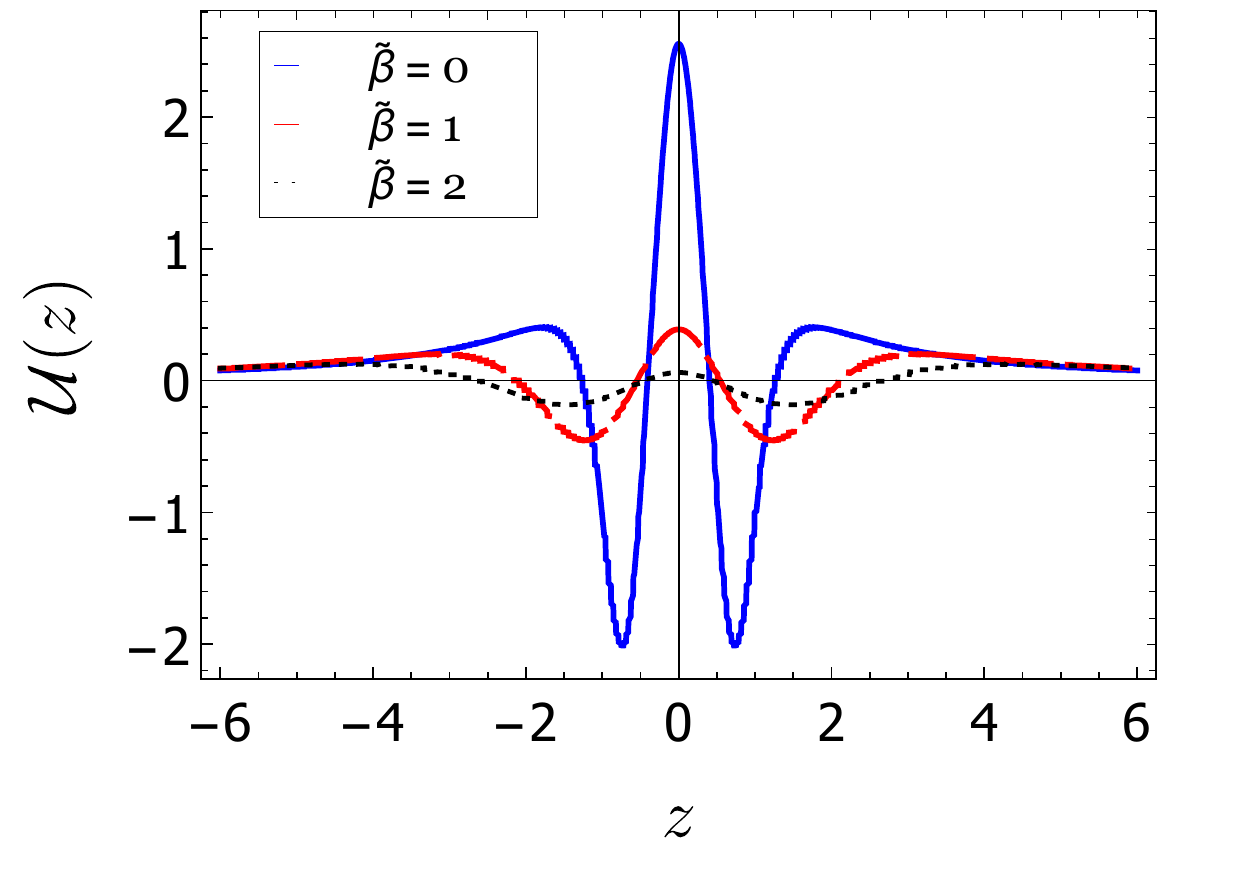}\\
\includegraphics[width=7cm,trim={0cm 0cm 0 0cm},clip]{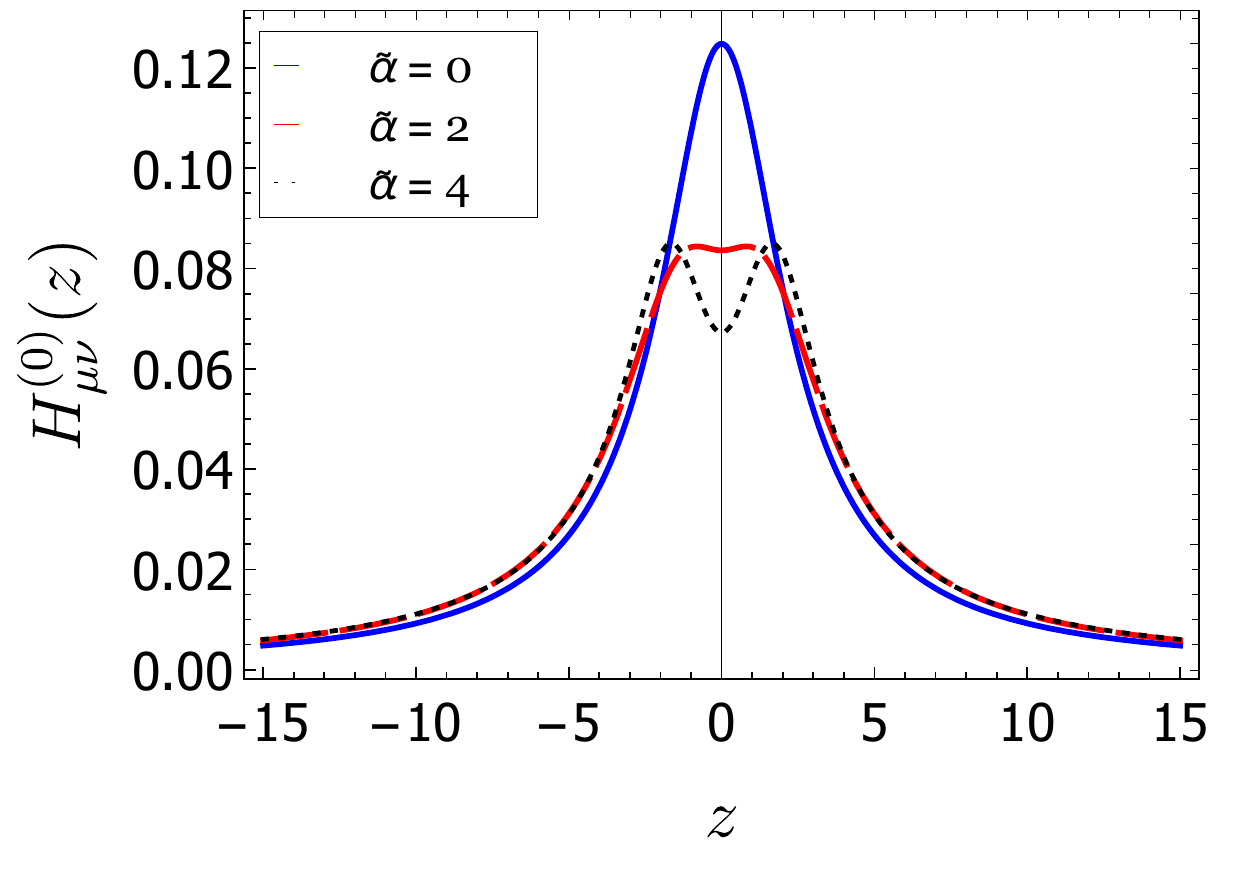}
\includegraphics[width=7cm,trim={0cm 0cm 0 0cm},clip]{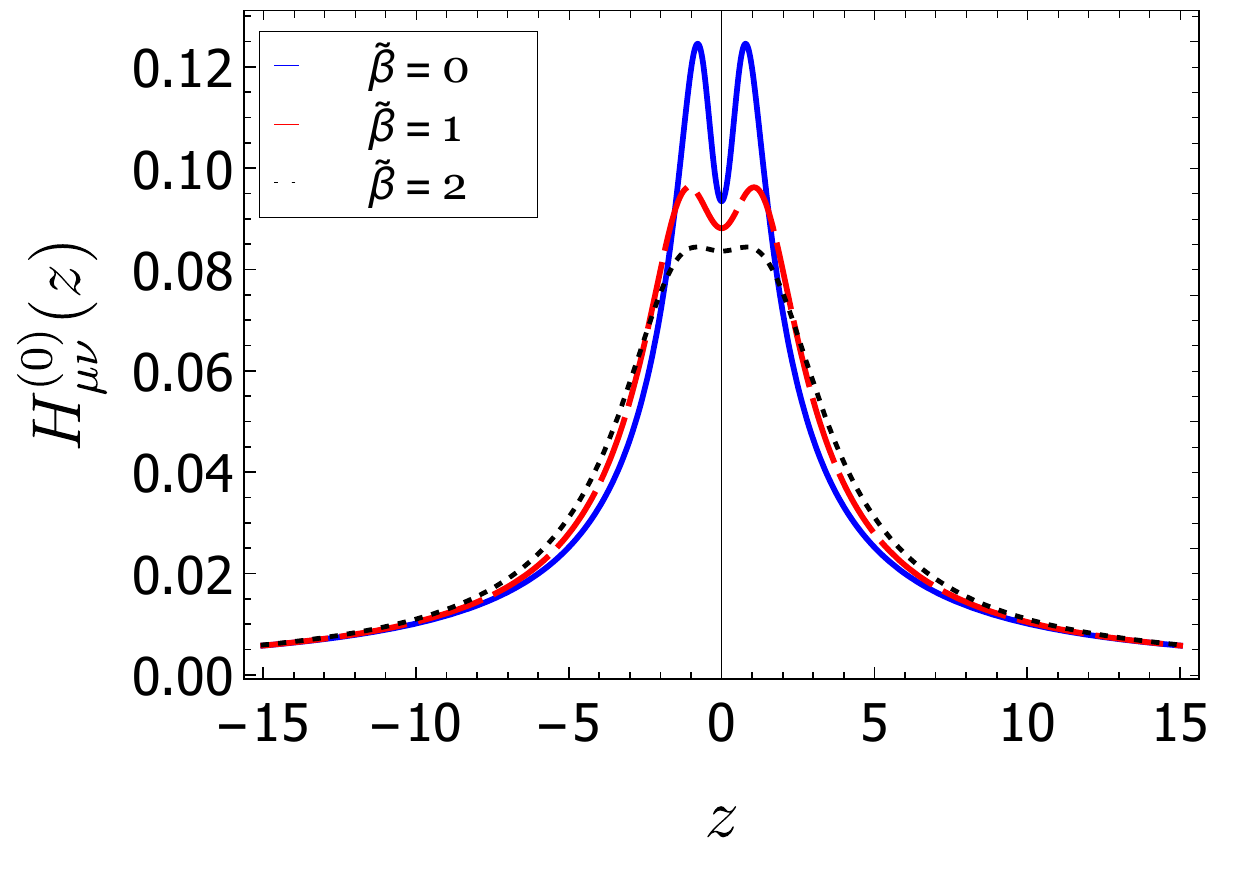}
\vspace{-0.3cm}
\caption{Stability potential for $\beta=2$ and $\alpha = \{0, 2, 4\}$ (top left panel), and for $\alpha=2$ and $\beta=\{0,1,2\}$ (top right panel). Graviton zero mode for the same values of $\alpha$ and $\beta$ (bottom panels). In both cases we use $\kappa=2$.}
\label{fig6}
\end{figure*}

Fig. \ref{fig6} shows the stability potential ${\cal U}$ and the zero mode $H_{\mu\nu}^{(0)}(z)$ for the special case where $f(R,T)=R+\beta T$ investigated in Sec. \ref{speccase}. We verify that for large enough values of $\alpha$, the behavior of the potential changes qualitatively from a global minimum to a local maximum at $y=0$. This transition is only visible for positive values of $\alpha$, while their negative counterparts simply increase the depth of the potential well at $y=0$. For even larger values of $\alpha$, the new maximum at $y=0$ may raise to positive values, thus inducing a second transition at which the potential well at $y=0$ becomes a potential barrier. When this happens, the zero mode suffers a split at $y=0$, which transitions from a global maximum to a local minimum. We thus conclude that positive contributions of the Gauss-Bonnet term can lead to the rise of interesting new behaviors of the stability potential and, consequently, to the development of an internal structure at the brane. On the other hand, a variation in the parameter $\beta$ maintains the qualitative behavior of both the stability potential and zero mode, while affecting solely the width and height of the peaks and droughts.

\begin{figure}[t!]
\centering
\includegraphics[width=7cm,trim={0cm 0cm 0 0cm},clip]{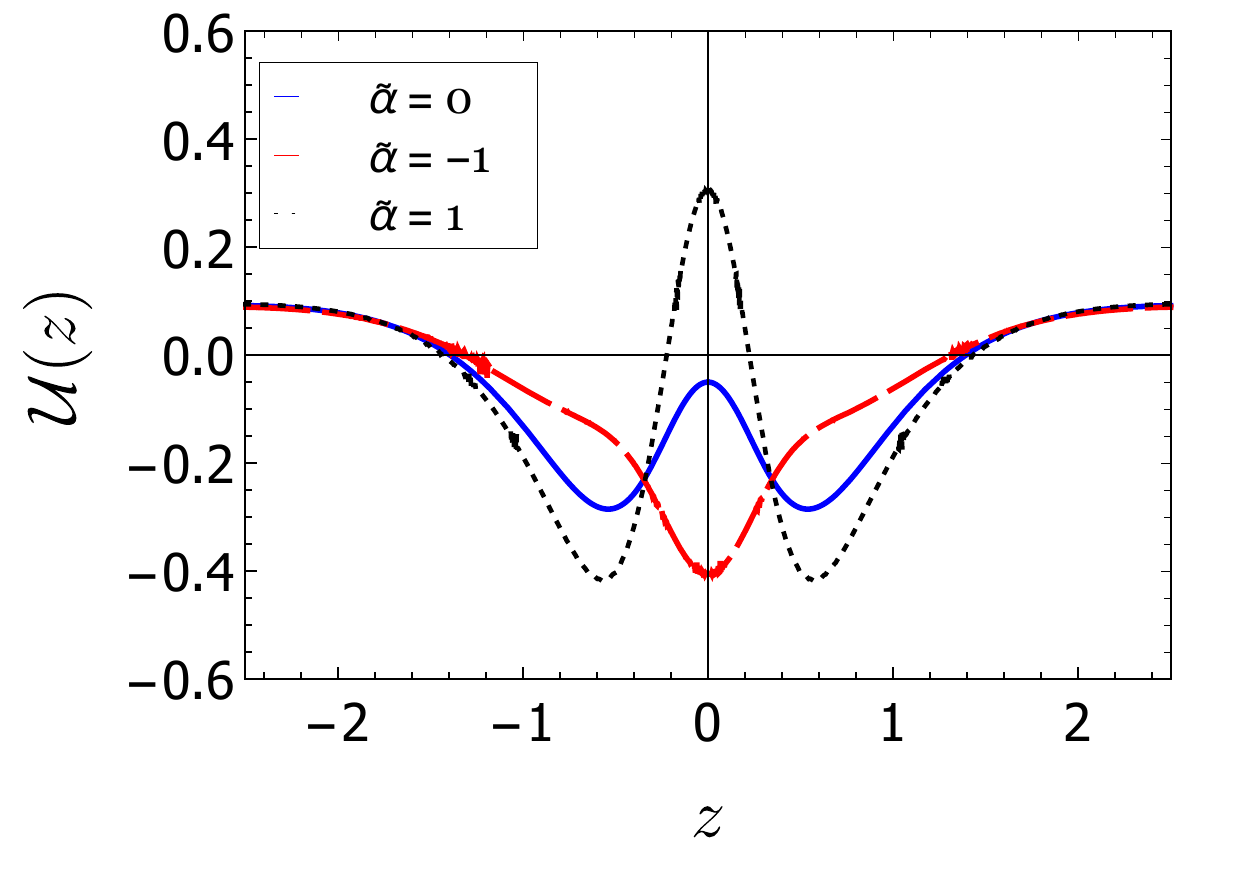}
\includegraphics[width=7cm,trim={0cm 0cm 0 0cm},clip]{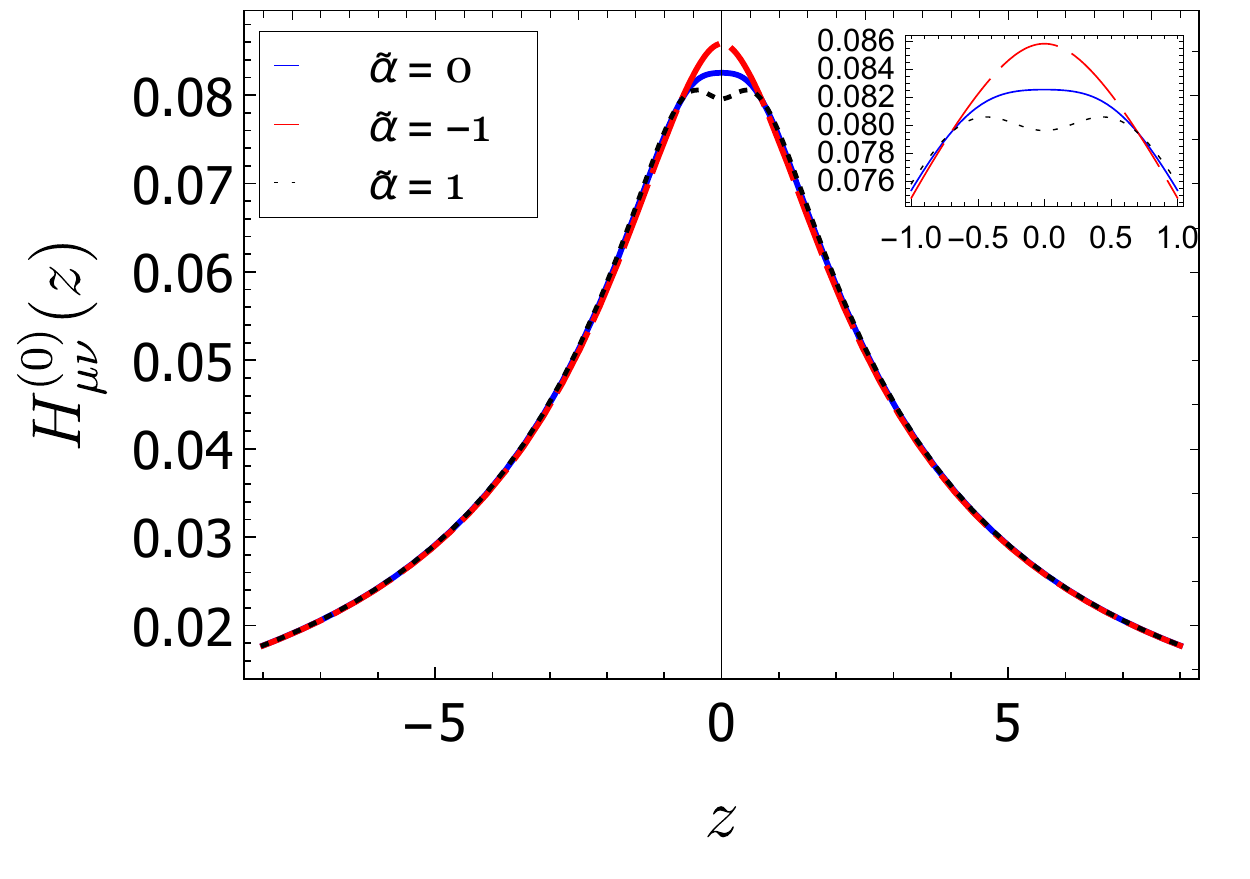}
\vspace{-0.3cm}
\caption{Stability potential for $\alpha = \{-1, 0, 1\}$ (top panel). Graviton zero mode for the same value of $\alpha$ (bottom panel). In both cases we use $\kappa=2$, $\psi(0)=5$, $\chi(0)=-1$, and $U(0)=0$.}
\label{fig7}
\end{figure}

As for the models investigated through the scalar-tensor representation, the qualitative behaviors of the stability potential and zero mode are roughly the same, i.e., the stability potential may feature either a global minimum at $y=0$, a local negative maximum at $y=0$, or a positive (eventually global) maximum at $y=0$, depending on the value of $\alpha$. In Fig. \ref{fig7} show the stability potential and the zero mode for the first model investigated in Sec. \ref{models}. Given the similarities of the results among the remaining models studied, we chose not to present those results explicitly. Unlike it happens for the $f\left(R,T\right)=R+\beta T$ case, negative values of $\alpha$ may induce non-trivial changes in the behavior of the potential, and consequently the structure of the zero mode, depending on the boundary conditions chosen for $\psi(0)$. Overall, larger values of $\alpha$ tend to increase the height of the maximum at $y=0$, while smaller (negative) values of $\alpha$ induce the opposite effect, decreasing the height of the maximum and eventually causing a transition to a global minimum, i.e., a single potential well. In accordance with the expected, the internal structure of the zero mode develops whenever the maximum at $y=0$ is positive, i.e., the potential well transitions into a potential barrier at $y=0$.

\section{Comments and conclusions}\label{coments}

In this paper, we investigate braneworld models in a generalized gravity theory by including a Gauss-Bonnet invariant together with a function of the trace of the stress-energy tensor and the Ricci scalar. We use the linear form $F(R,G,T)=f(R,T)+\alpha G$, where $\alpha$ is a real parameter that controls the contribution of the Gauss-Bonnet term. In general, we found the conditions that guarantee the stability of the model under tensor perturbations and described the zero mode. Furthermore, we obtain solutions for the specific case where $f(R,T)=R+\beta T$ and for the general case through the scalar-tensor formalism.

Firstly, we analyzed the particular case where $f(R,T)=R+\beta T$ and obtained analytical solutions for the source field and warp factor. We verified that the brane behaves in the desired way, preserving the regularity of the Kretschmann scalar and achieving the correct asymptotic behavior. We verified that the Gauss-Bonnet term may induce a split in the stability potential if its contribution to the action is large enough, causing a transition of the central point from a minimum to a maximum in $y=0$. If $\alpha$ is large enough, which seems to happen at $\alpha\gtrsim 1$, the zero mode exhibits an internal structure which increases with $\alpha$.

We also investigated the general case of the $f(R,T)-$brane through the scalar-tensor representation. We implemented the necessary mathematical description to use the scalar-tensor representation and used the first-order formalism to simplify the equations. To exemplify what was done, we investigated some simple models that allow for kink-like topological structures. We found that the Gauss-Bonnet term introduces changes only in the asymptotic behavior of the solutions while maintaining the behavior at the origin unaltered. This feature allows one to find qualitatively different solutions for different values of $\alpha$, depending on the boundary conditions chosen at $y=0$. The quantities associated with stability, specifically the stability potential and the zero mode, have their central peak modified by the $\alpha$ parameter. In general, the peak at $y=0$ can be controlled by adjusting the contribution of the Gauss-Bonnet invariant, which also allows one to transition between different qualitative behaviors e.g. a single potential well, a double potential well, and a potential barrier, the latter inducing internal structure in the graviton zero mode.

Therefore, we verified that the addition of a quadratic invariant in the form of the Gauss-Bonnet invariant can qualitatively modify the profile of the fields in the brane model investigated in the scalar-tensor representation. This investigation is interesting as it contributes to a better understanding of how modified gravity models work in higher dimensions. On the other hand, it is also interesting to include other scalar invariants in the study of braneworld models. In particular, the so-called cubic gravity has piqued our interest and we believe that it can also be incorporated into the scalar-tensor formalism of the $f(R,T)-$brane. These and other investigations are currently under consideration and we look forward to reporting on them soon.

\begin{acknowledgments}

DB would like to thank financial support from CNPq, grant No. 303469/2019-6. DB and ASL also thank Paraiba State Research Foundation, FAPESQ-PB, grant No. 0015/2019, for partial financial support. JLR acknowledges the European Regional Development Fund and the programme Mobilitas Pluss for financial support through Project No.~MOBJD647, and project No.~2021/43/P/ST2/02141 co-funded by the Polish National Science Centre and the European Union Framework Programme for Research and Innovation Horizon 2020 under the Marie Sklodowska-Curie grant agreement No. 94533.

\end{acknowledgments}


\end{document}